\documentclass[12pt,preprint]{aastex}
\usepackage{amssymb}
\usepackage{epsfig}
\usepackage{natbib}
\usepackage{graphicx}
\usepackage{color}
\usepackage{tabularx}
\usepackage{epsfig}
\usepackage{rotating}
\usepackage{multirow}
\usepackage{subfigure}

\newcommand       \mum        {\,{\rm \mu m}}

\newcommand       \Ks           {{K_{\rm s}}}

\newcommand       \simali       {\,{\sim}}
\newcommand       \magni        {\,{\rm mag}}
\newcommand       \Angstrom     {\,{\rm \AA}}

\newcommand       \Teff         {T_{\rm eff}}
\newcommand       \K            {\,{\rm K}}

\newcommand       \Cll       {C_{\lambda_1\lambda_2}}
\newcommand       \CBV      {C_{\rm BV}}

\newcommand       \CVJ       {C_{\rm VJ}}

\newcommand       \CKWa       {C_{\rm K_SW1}}
\newcommand       \CKWb       {C_{\rm K_SW2}}

\newcommand       \CBVint      {C^0_{\rm BV}}

\newcommand       \CVJint       {C^0_{\rm VJ}}

\newcommand       \CgV       {C_{\rm gV}}
\newcommand       \CVr       {C_{\rm Vr}}
\newcommand       \CVi       {C_{\rm Vi}}
\newcommand       \CVH       {C_{\rm VH}}
\newcommand       \CVK       {C_{\rm VK_S}}

\newcommand       \CgVint       {C^0_{\rm gV}}
\newcommand       \CVrint       {C^0_{\rm Vr}}
\newcommand       \CViint       {C^0_{\rm Vi}}
\newcommand       \CVHint       {C^0_{\rm VH}}
\newcommand       \CVKint       {C^0_{\rm VK}}
\newcommand       \CVWaint       {C^0_{\rm VW1}}
\newcommand       \CVWbint       {C^0_{\rm VW2}}
\newcommand       \EBV       {E_{\rm BV}}
\newcommand       \EgV       {E_{\rm gV}}
\newcommand       \EVr        {E_{\rm Vr}}
\newcommand       \EVi         {E_{\rm Vi}}
\newcommand       \EVH       {E_{\rm VH}}
\newcommand       \EVK       {E_{\rm VK_S}}
\newcommand       \EVWa       {E_{\rm VW1}}
\newcommand       \EVWb       {E_{\rm VW2}}

\newcommand       \EVJ       {E_{\rm VJ}}

\newcommand       \EVl       {E_{\rm V\lambda_x}}

\newcommand{\Alx}{A_{\lambda_x}}
\newcommand{\AV}{A_{V}}

\newcommand{\AKs}{A_{\rm K_S}}

\newcommand{\RV}{R_{V}}
\newcommand{\Rv}{R_{V}}

\shorttitle{The Extinction Law of the $l165^{\circ}$ Sightline
         in the Galactic Plane}
\shortauthors{Wang et al.}

\begin{document}

\title{
%------------- enable for labelling preprint ---------------------------
% \vspace*{-2.0em}
%  {\normalsize\rm submitted to {\it The Astrophysical Journal}}\\
% \vspace*{1.0em}
%---------------------------------------------------------------
The Optical -- Mid-infrared
Extinction Law of the $l=165^{\circ}$  Sightline in the Galactic Plane:
\\Diversity of Extinction Law in the Diffuse Interstellar Medium
     }

\author{Shu Wang\altaffilmark{1,2},
             B.W.~Jiang\altaffilmark{2}, 
             He Zhao\altaffilmark{2}, 
             Xiaodian Chen\altaffilmark{3}, and
             Richard de Grijs\altaffilmark{1,4}}
\altaffiltext{1}{Kavli Institute for Astronomy and Astrophysics,
                 Peking University,
                 Beijing 100871, China;
                 {\sf shuwang@pku.edu.cn}
                 }
\altaffiltext{2}{Department of Astronomy,
                 Beijing Normal University,
                 Beijing 100875, China;
                 {\sf bjiang@bnu.edu.cn}
                 }
\altaffiltext{3}{Key Laboratory for Optical Astronomy, 
                 National Astronomical Observatories, 
                 Chinese Academy of Sciences, 
                 Beijing 100012, China
                 }
\altaffiltext{4}{International Space Science Institute--Beijing,
                 Beijing 100190, China
                 }

\begin{abstract}

Understanding the effects of dust extinction is important to properly
interpret observations. The optical total-to-selective extinction
ratio, $\RV = \AV/E(B-V)$, is widely used to describe extinction
variations in ultraviolet and optical bands. Since the $\RV=3.1$
extinction curve adequately represents the average extinction law of
diffuse regions in the Milky Way, it is commonly used to correct
observational measurements along sightlines toward diffuse regions in
the interstellar medium. However, the $\Rv$ value may vary even along
different diffuse interstellar medium sightlines. In this paper, we
investigate the optical--mid-infrared (mid-IR) extinction law toward a
very diffuse region at $l = 165^{\circ}$ in the Galactic plane, which
was selected based on a CO emission map. Adopting red clump stars as
extinction tracers, we determine the optical-to-mid-IR extinction law
for our diffuse region in the two APASS bands ($B, V$), the three
XSTPS-GAC bands ($g, r, i$), the three {\it 2MASS} bands ($J, H,
\Ks$), and the two {\it WISE} bands ($W1, W2$). Specifically, 18 red
clump stars were selected from the APOGEE--RC catalog based on
spectroscopic data in order to explore the diversity of the extinction
law. We find that the optical extinction curves exhibit appreciable
diversity. The corresponding $\Rv$ ranges from 1.7 to 3.8, while the
mean $\Rv$ value of 2.8 is consistent with the widely adopted average
value of 3.1 for Galactic diffuse clouds. There is no apparent
correlation between $\Rv$ value and color excess $E(B-V)$ in the range
of interest, from 0.2 to 0.6 mag, or with specific visual extinction
per kiloparsec, $\AV/d$.

\end{abstract}


\section{Introduction}\label{intro}
%

%extinction law
The continuous interstellar extinction, the `extinction law' along
each sightline, or the variation of extinction with wavelength
$\lambda$ are usually expressed as a ratio of color excesses---such as
$E(\lambda-V)/E(B-V)$---or of the absolute extinction (such as
$A_\lambda/\AV$), where adoption of the $B$ and $V$ bands as reference
bands is a convention from the optical era. The information about the
extinction law is independent as to how the law is expressed.
%UV/optical
The ultraviolet (UV)/optical extinction at $\lambda < 0.9\mum$ is
known to vary significantly between sightlines. Cardelli et
al.\ (1989; hereafter CCM89) explored the extinction laws in various
environments, including in diffuse regions, molecular clouds, and
H{\sc ii} regions, over the available wavelength ranges. CCM89 used
the optical total-to-selective extinction ratio $\Rv=\AV/E(B-V)$ to
describe extinction variations in UV/optical bands. Sightlines towards
the low-density interstellar medium (ISM) are usually characterized by
rather small $\Rv$ values, as low as $R_V \sim 2.1$ (sightline towards HD210121, Welty \& Fowler 1992), 
with an average of $R_V = 3.1$ (see Draine 2003; Schlafly \& Finkbeiner 2011).
Sightlines penetrating into dense clouds usually show rather high values of $\Rv$, such as the Ophiuchus or Taurus molecular clouds with $ 4 < \Rv < 6$ (see Mathis 1990). More recently, Schlafly et al.\ (2016) measured optical--infrared (IR)  reddening values to 37,000 stars in the Galactic disk, with fewer than 1\% of sightlines having $\Rv > 4$. 

%near-IR
As the extinction law exhibits significant differences in various
environments at UV/optical wavelengths, one might expect corresponding
variations at longer, IR wavelengths. Previous studies have
found that the near-IR extinction, within the wavelength range
$0.9\mum < \lambda < 3\mum$, follows a power law defined by
$A_{\lambda}\propto{\lambda^{-\alpha}}$, with the index spanning a
small range of $1.61 < \alpha < 1.80$ (Draine 2003). Starting from the
current century, newly derived values of $\alpha$ have become
systematically larger, mostly $\alpha >2.0$ (Wang \&
Jiang\ 2014). Wang \& Jiang\ (2014) re-investigated the near-IR
extinction law by using a sample of K-type giants selected from the
APOGEE spectroscopic survey. They confirmed that the near-IR
extinction law is universal, with $E(J-H)/E(J-\Ks)=0.64$,
corresponding to $\alpha=1.95$.
%mid-IR
Meanwhile, the mid-IR ($3\mum <\lambda< 8\mum$) extinction law seems
flat in both diffuse and dense environments, as suggested by studies
along numerous sightlines, including toward the Galactic Center (Lutz
1999; Nishiyama et al.\ 2009), the Galactic plane (Indebetouw et
al.\ 2005; Jiang et al.\ 2006; Gao et al.\ 2009), and nearby
star-forming regions (Flaherty et al.\ 2007). Wang et al.\ (2013)
investigated the mid-IR extinction law and its variation in the
Coalsack nebula. They found that the mid-IR extinction curves are all
flat and the relative extinction $A_\lambda/\AKs$ decreases from
diffuse to dense environments in the four {\it Spitzer} IRAC bands.
In addition, there is some evidence that the mid-IR extinction law may
vary. Gao et al.\ (2009) claimed that the 3--8$\mum$ extinction law
may vary with Galactic longitude (see also Zasowski et
al.\ 2009). However, their results disagree about the actual
variations in the extinction law, although both studies are based on
very similar data and methods. 
Xue et al.\ (2016) obtained precise average mid-IR extinction law 
and found no apparent variation with the extinction depth. 
Thus, the IR extinction law may be universal, and its variation, if any, is small.

%this work aims at...
Since the CCM89 $\RV=3.1$ extinction curve adequately represents the
average extinction law of diffuse regions, it is commonly used to
correct observations for the effects of interstellar extinction along
diffuse ISM sightlines. However, a given value of $\Rv$ may not be
able to reflect the true interstellar environment along some lines of
sight. For example, the star Cyg OB2 12, the 12th brightest member of
the Cygnus OB2 association, is located behind a dense cloud (Mathis
1990) or a pile-up of diffuse molecular clouds along the line of sight
(Snow \& McCall 2006), but it has $\Rv = 2.65$ (Clark et al.\ 2012;
$\AV = 10.18$ mag) or $\Rv = 3.04$ (Torres-Dodgen et al.\ 1991; $\AV
=10.20$ mag), a value appropriate for the diffuse ISM. Moreover, for a
true sightline, there exists apparent deviation from the CCM89
analytical extinction curve calculated for a given value of $\Rv$
(Mathis 1990). The extinction toward HD 210121, located behind the
core of a molecular cloud (De\'sert et al. 1988; de Vries \& van
Dishoeck 1988), can be best fitted by the CCM89 $\Rv$ = 2.1 curve, but
it shows a significantly lower bump at 2175 $\Angstrom$ and a much
steeper rise in the far-UV compared with the average behavior for the
same value of $\Rv$ (Larson et al.\ 2000). The present work aims at
revealing the diversity of the extinction law in diffuse regions by
carefully examining a very diffuse sightline covering an area of four
square degrees.
To achieve this goal, we first explore interstellar environments in
the Galactic plane and search for a diffuse region (Section 2). Then,
we investigate the extinction laws characteristic of the diffuse
region by means of red clump (RC) stars (Sections 3 and 4). Finally,
we analyze the diversity of the extinction law (Section 5).

\section{The Diffuse Region: G$l165.0^{\circ}+0.0$} \label{diffuse}
\subsection{Selection criteria}

Three criteria can independently be used to characterize the ISM:
Visual extinction, IR dust emission, and CO line intensity.
%Av
The extinction $A_\lambda$ depends on the dust column density, $N_{\rm
  d}$, and the optical properties of the dust: $A_\lambda =
1.086\,N_{\rm d}\,C_{\rm ext}(a,\lambda)$, where $C_{\rm
  ext}(a,\lambda)$ is the extinction cross-section of the dust with a
typical size $a$ at a wavelength $\lambda$. Hence, a high $A_\lambda$
implies a dense cloud or a pile-up of many diffuse clouds along the
line of sight.
%emission
The dust-emission intensity $I_\lambda$ is proportional to the dust
column density, the absorption cross-section, and the specific
emission intensity of the dust: $I_\lambda \propto N_{\rm d}\,C_{\rm
  abs}(a,\lambda)\,B_\lambda(T)$, where $B_\lambda(T)$ is the Planck
function at the dust temperature $T$ and wavelength $\lambda$.
%I(CO)
The high intensity of the CO emission line is often used to indicate
dense interstellar environments. Generally speaking, with the
intensity of the CO emission line, $I$(CO), the total mass of the
molecular gas can be derived using an empirical CO-to-H$_2$ conversion
factor $X_{\rm CO} \equiv N_{\rm H_2}/I_{\rm CO}$. If the gas is well
mixed with the dust, with a constant gas-to-dust ratio, the total dust
mass can be derived. Thus, the intensity of the CO emission line is
also proportional to the dust column density. Indeed, Zasowski et
al. (2009) used the $^{\rm 13}$CO ($J$=1--0) line to trace dense
interstellar clouds. Wang et al.\ (2013) used these criteria to
distinguish complex environments in the Coalsack nebula region to
investigate the variation of the mid-IR extinction law.

\subsection{The G$l165.0^{\circ}+0.0$ Region} \label{diffuse}

\begin{figure}[h!]
\centering
\vspace{-0.2in}
\includegraphics[angle=0,width=6.7in]{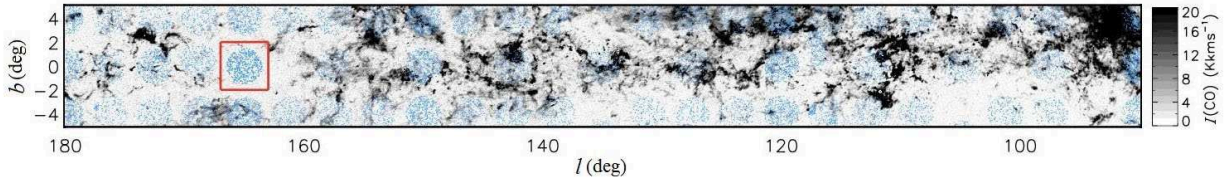}
\vspace{-0.5in}
\caption{\footnotesize
               \label{fig:co}
          {\it Planck} velocity-integrated CO (1--0) emission
          intensity contours of the Galactic plane region $\mid b \mid
          \le 5^{\circ}$. Blue dots: APOGEE giants; red square:
          selected G$l165^{\circ}\pm0$ region.  }
\end{figure}

As described in the previous section, the CO (1--0) line intensity
$I$(CO) is often used to characterize dense interstellar environments.
Integrated CO line intensity contours reveal complex Galactic
structures. Guided by this, we checked the CO emission intensity maps
of the Galactic plane to find candidate diffuse ISM regions.
%CO data
A wealth of CO line intensity databases exists for the Galactic plane
and some well-known molecular clouds. The Milky Way's CO emission map
of Dame et al.\ (2001) covers the entire Galactic plane at Galactic
latitudes $\mid b \mid \le 30^{\circ}$ with an effective angular
resolution of 0.5$^{\circ}$. The European Space Agency's {\it Planck}
satellite observed the sky in nine bands covering frequencies of
30--857 GHz with high sensitivity and high spatial resolution. {\it
  Planck} CO maps have been extracted from the {\it Planck} HFI
data. In this work, we adopt the {\it Planck} Type 3 CO (1--0) map
because of its high resolution and sensitivity (Planck Collaboration
XIII et al.\ 2014). The angular resolution is $5.5'$, and the standard
deviation is 0.16 K km s$^{-1}$ at an angular resolution of $15'$. 
For comparison, the CO survey of Dame (2011) has a typical uncertainty of
0.6 K km s$^{-1}$. Figure~\ref{fig:co} displays integrated CO line
intensity contours for $90^{\circ} < l < 180^{\circ}$, $\mid b \mid \le 5^{\circ}$, 
where molecular clouds stand out clearly because of their intense CO line emission, 
while some regions are diffuse with low CO line intensities. 
Our candidate regions are restricted to the Galactic plane because 
the stars in the Galactic halo are usually metal-poor.  
We plan to use RC stars as the tracers of the interstellar extinction. 
The intrinsic colors of RC stars are needed to be determined (Section 4). 
The optical intrinsic color of RC stars relates to their metallicity 
(e.g. Sarajedini 1999; Girardi \& Salaris 2001; Nataf et al.\ 2010). 
Therefore, if we were to include stars in the Galactic halo, 
we should also consider the effect of metallicity in deriving the intrinsic color for RC stars. 
For this reason, we limit our candidate diffuse regions in the Galactic plane.
%APOGEE
In addition, RC stars will be selected based on the APOGEE survey, 
the diffuse region targeted must have been observed by it.  
The APOGEE survey targeted more than 100,000 red-giant
stars selected from the {\it 2MASS} database (Skrutskie et al.\ 2006),
for which accurate stellar parameters were derived.
%find diffuse
However, APOGEE is not an all-sky survey. Therefore, the diffuse
region targeted here was chosen by overlaying the APOGEE stars on the
integrated CO emission intensity map. In Figure~\ref{fig:co}, the blue
dots are giants from APOGEE.

%figure for CO+APOGEE

%l165
Based on Figure~\ref{fig:co}, we selected one candidate diffuse
region, `$l165^{\circ}$,' centered on ($l = 165.0^{\circ}, b =
0.0^{\circ}$). It covers $4^{\circ} \times 4^{\circ}$ on the sky, and
the mean $I$(CO) is about 2 K km s$^{-1}$; the region is indicated by
the red square. The other two factors, the visual extinction and the
dust IR emission intensity, are also used in selecting the diffuse
region (see for more details Section 1). The $l165^{\circ}$ region
shows essentially very little extinction with $\AV\sim1.0 \magni$ in
the visual extinction maps of Dobashi et al.\ (2005) and from Chen et
al.\ (2015). This amount of extinction matches very well the
characteristic extinction of diffuse clouds, i.e., the total visual
extinction $\AV$ is $\sim$0--1 mag (Snow \& McCall 2006; Draine
2011). In addition, the {\it Spitzer}/MIPS 24$\mum$ image shows no
detectable 24$\mum$ emission for $l165^{\circ}$. Therefore, the CO
line intensity, visual extinction, and dust emission all indicate that
the $l165^{\circ}$ region is very diffuse.

\section{Data and Tracers}

The extinction is most reliably determined by comparing
spectrophotometry of two stars (one with negligible foreground dust,
the other heavily reddened) of the same spectral class under the
assumption that the dust extinction decreases to zero at very long
wavelengths (Draine 2003). Lutz (1996) probed the extinction law
toward the Galactic Center between 2.5 and $19\mum$ by comparing the
observed and expected intensity ratios of the hydrogen recombination
lines. In addition, the color-excess method is widely applied to
photometric data; it can probe more deeply than the spectrum-pair
method. Hence, most IR extinction studies are performed using the
color-excess method. In brief, this statistical method computes the
ratio of two color excesses for a group of tracers with homogeneous
intrinsic color indices. Red-giant stars (RGs) and RC stars are
appropriate tracers. The advantages of using these tracers are that
they are bright and numerous, and they can be selected based on
near-IR color--magnitude diagrams (CMDs). Their disadvantage is that
contamination by other types of stars is unavoidable (Wang \& Jiang
2014). Recently, Wang \& Jiang (2014) and Xue et al. (2016) adopted a
new method that combines photometry and spectroscopy to derive
accurate stellar extinction values. In this work, we will follow their
method to investigate the optical-to-mid-IR extinction law of the
$l165^{\circ}$ diffuse region. In essence, the intrinsic color index
is calculated based on the stellar parameters, and the color excess is
subsequently derived.

\subsection{Data}

Broad-band photometric data from APASS, XSTPS-GAC, {\it 2MASS}, and
\emph{WISE} are used to derive the observed colors, while the stellar
spectroscopic data set from APOGEE is used to determine the intrinsic
colors.

\subsubsection{Optical to Infrared Photometric Data:
APASS, XSTPS-GAC, {\it 2MASS}, WISE Surveys}
%

%APASS DR9
The American Association of Variable Star Observers (AAVSO)
Photometric All-Sky Survey (APASS) is conducted in five filters:
Landolt $B$ and $V$ and Sloan $g', r'$, and $i'$. 
The reliable magnitude range in the $V$ band runs 
from $V = 7 \magni$ to 17$\magni$ (Henden \& Munari 2014). 
The latest, DR9 catalog contains photometry for approximately 
62 million objects, covering about 99\% of the sky (Henden et al.\ 2016). 
Munari et al.\ (2014) investigated the external accuracy of 
APASS photometry, based on secondary Landolt and Sloan photometric 
standard stars, and on a large body of literature data on field and cluster stars. 
They confirmed that the APASS photometry did not show any offsets or trends.
We obtained the $B$ and $V$ data from the APASS/DR9.

%Xuyi
Xuyi 1.04/1.20 m Schmidt Telescope Photometric Survey of the Galactic
Anticenter (XSTPS-GAC) consists of two parts, one centered in the
Galactic Anticenter area covering $140^{\circ} < l < 240^{\circ}$ and
$-60^{\circ} < b< 40^{\circ}$, the other covering the M31/M33 area
(for more details, see Liu et al.\ 2014; Zhang et al.\ 2014). The
XSTPS-GAC photometric catalog provides $g$-, $r$-, and $i$-band data
for more than 100 million stars; the passbands are the same as for the
Sloan Digital Sky Survey (SDSS)\footnote{Note that the SDSS filters
  are called $g', r'$, and $i'$. However, in the XSTPS-GAC catalog,
  $g, r$, and $i$ are used to refer to these same filters}. The
limiting magnitude is about 19$\magni$ in the $r$ band ($\sim10\sigma$), 
with an astrometric accuracy of $\sim0.1$ arcsec (Liu et al.\ 2014). 
Chen et al.\ (2014) and Liu et al.\ (2014) pointed out that 
flux calibration with respect to the SDSS photometry produces 
a photometric accuracy of better than 2\% for a single frame 
and $\sim$2--3\% for the whole observation area.
We take the $g, r$, and $i$ photometric data from
the sample of Chen et al.\ (2014), who used it to determine the
three-dimensional extinction map of the Galactic Anticenter.

%2MASS
The Two Micron All Sky Survey (2MASS) is a near-IR ground-based
whole-sky survey using two 1.3 m aperture telescopes (Skrutskie et
al.\ 1997). Over 470 million sources in its point-source catalog
provide measurements in the $J$, $H$, and $\Ks$ bands.

%WISE
The Wide-Field Infrared Survey Explorer ({\it WISE}) survey is a
full-sky mid-IR survey with a 40 cm space-borne telescope (Wright et
al.\ 2010). It mapped the sky in the $W1, W2, W3$, and $W4$ bands
(with central wavelengths of, respectively, 3.4, 4.6, 12, and
22$\mum$) and yielded a source catalog of over 563 million objects
with 5$\sigma$ photometric sensitivities of about 0.068, 0.098, 0.86,
and 5.4 mJy in $W1, W2, W3$, and $W4$, respectively, in unconfused
regions along the ecliptic plane. Because the sensitivities of the
$W3$ and $W4$ bands are relatively low, we only consider the $W1$ and
$W2$ bands here. The {\it WISE} photometric data are taken 
from the sample of Chen et al.\ (2014), who adopted the {\it WISE} All-Sky source catalog\footnote{For our sample (18 RC stars, section 3.2), the photometric data 
adopted from the {\it WISE} All-Sky source catalog or the AllWISE source catalog 
will introduce magnitude differences. However, the differences 
in the $W1$ and $W2$ bands are less than 0.2\% and 0.1\%, respectively.
In addition, we did not find any systematic trends.}.

\subsubsection{Spectroscopic Data: The SDSS/APOGEE Survey}

%APOGEE
APOGEE is a near-IR $H$-band (1.51--1.70$\mum$), high-resolution
($R\sim 22,500$) spectroscopic survey. Part of SDSS-III DR12, it
includes all data obtained from 2008 August to 2014 June\footnote{The
  latest data release is DR13, containing observations through 2015
  July. Compared with DR12, there are about 1000 additional sources in
  DR13. Since DR13 has been re-calibrated, the DR13 stellar parameters
  are slightly different compared with those contained in DR12,
  especially the stellar surface gravities. Since the APOGEE RC
  catalog is based on the DR12 parameters, we adopted the data from
  DR12.} (Alam et al.\ 2015). The APOGEE Stellar Parameter and
Chemical Abundances Pipeline (ASPCAP) extracts stellar parameters,
including effective temperatures $\Teff$, surface gravities $\log g$,
and detailed elemental abundances such as metallicities [Fe/H]
(Holtzman et al.\ 2015). The uncertainties are typically 50--100 K in
$\Teff$, 0.2 dex in $\log g$, and 0.03--0.08 dex in [Fe/H]
(M\'esz\'aros et al.\ 2013). It is a good tool to investigate the
composition and dynamics of stars in the Galaxy.
%APOGEE DR12 RC catalog
The APOGEE data release also includes the APOGEE red-clump
(APOGEE--RC) catalog from DR11 and DR12. The stellar parameters in the
APOGE--RC catalog are based on APOGEE data and calibrated using
stellar evolution models and asteroseismology data. RC stars are
selected by their position in
color--metallicity--surface-gravity--effective-temperature space (Bovy
et al.\ 2014). The APOGEE--RC DR12 catalog contains about 20,000
likely RC stars with an estimated contamination of less than 3.5\%
(Bovy et al.\ 2014).

%Summary
By cross-matching the photometric and spectroscopic catalogs, we have
constructed a multiband stellar sample. To summarize, in total
ten-band optical-to-IR (i.e., $B, V, g, r, i$, $J$, $H$, $\Ks$, $W1$, and
$W2$) photometric data have been collected from the APASS, XSTPS-GAC,
{\it 2MASS}, and {\it WISE} survey programs. The stellar parameters
$\Teff$, $\log g$, and [Fe/H] were extracted from the APOGEE catalog.

\subsection{Tracers}

RGs and RC stars are frequently used as IR interstellar extinction
tracers.
%Red giants
RGs with a small scatter in the IR intrinsic color index
($C_{J\Ks}^0$; Gao et al.\ 2009; Wang et al.\ 2013) are usually
selected based on mid-IR color restrictions: $[3.6]-[4.5]<0.6$\,mag
and $[5.8]-[8.0]<0.2$\,mag (Flaherty et al.\ 2007). Although these
criteria could effectively remove sources with intrinsic IR excesses,
some asymptotic giant-branch (AGB) stars that suffer from
circumstellar extinction may contaminate the RG sample.
%Red Clumps
RC stars are a group of K2III-type stars in the core-helium-burning
stage. Their absolute magnitude is around $M_{K} = -1.61 \pm 0.03
\magni$ (Alves 2000), and their near-IR intrinsic color index is
$0.65\leq C_{J\Ks}^0 \leq0.75$ mag \citep{Wainscoat92, GF14, WJ14}.
Because of the constant IR luminosity and very small scatter in
$J-\Ks$, their distribution in the near-IR $(J-\Ks)$ versus $\Ks$ CMD
forms a narrow strip, which is commonly adopted to identify RCs.
However, the observed $(J-\Ks)$ color ($C_{J\Ks}$) depends only on the
interstellar extinction, while $\Ks$ magnitudes depend on both
interstellar extinction and distance, leading to a large dispersion of
RC stars in the CMD. Therefore, selection of RC stars from the CMD may
include some dwarf stars, specifically a fraction of $\simali$
2.5--5\% for $\Ks<12.5\magni$, and up to $\simali$ 10--40\% for
$13\magni<\Ks<14\magni$ \citep{LC02, CL07}. In addition, selection of
the RC strip in the CMD is not universal for all sightlines, and it is
also somewhat subjective on the basis of empirical and visual
inspection.

%Our sample
To avoid these uncertainties and contamination, we obtained a
homogeneous RC sample with the current-best available quality from a
combination of photometric and spectroscopic data. First, the
photometric quality must be $\sigma < 0.1$ mag in the $B$ and $V$
bands and $\sigma < 0.05$ mag in the $g, r, i, J, H, \Ks, W1$, and
$W2$ bands. As the metallicity [Fe/H] would affect the intrinsic color
at short wavelengths, we limit the [Fe/H] of giants ($\log g \le 3.0$)
to [Fe/H]$ > -0.5$ dex. Next, likely RC candidates were selected based
on their clumping in the $\Teff$--$\log g$ contour map resulting from
the entire APOGEE DR12 catalog, in the ranges $4550\K \le \Teff \le
5050\K$ and $2.3 \le \log g \le 3.0$.
In addition, the selected candidates were cross-matched with the
APOGEE--RC catalog. In fact, not all RC candidates are included in the
APOGEE--RC catalog. Therefore, our final, homogeneous RC sample
contains those RC candidates that are included in the APOGEE--RC
catalog.
For the diffuse $l165^{\circ}$ region, there are only 18 RC stars with
the full ten-band data of high quality. Their names, locations,
stellar parameters, and the $B, V$, and $J$-band photometric data are
listed in Table~\ref{tab:sources}, sorted by increasing $\Teff$.

\begin{table}[h!]
\begin{center}
\footnotesize
\caption{\label{tab:sources}
                Eighteen RCs in the $l165^{\circ}$ region
                selected for our extinction study}
\vspace{0.1in}
\begin{tabular}{ccccccccccc}
\hline \hline
No. & Name & $l$ & $b$ & $\Teff$ & $\log g$ & [Fe/H]
& $V$ & $B-V$ & $V-J$ \\
&  & ($^\circ$) & ($^\circ$) & (K) &   & 
& (mag) &  (mag) & (mag) \\
\hline
1& 2M05114361+4105251	& 166.05 & 0.97   & 4691.40 & 2.55 & 0.26 & 15.45 & 1.44 & 2.77\\ %1
2 &2M04564963+4125154	& 164.09 & $-$1.05  & 4844.28 & 2.45 & $-$0.19 & 14.61 & 1.53 & 2.83\\%2
3 &2M05113045+4121184 	& 165.82 & 1.10   & 4867.70 & 2.66 & $-$0.04 & 14.60 & 1.47	 & 2.41\\%3
4 &2M05103457+4157164	& 165.23 & 1.31   & 4878.76 & 2.69 & $-$0.45 & 13.99 & 1.24	 & 2.19\\%4
5 &2M05063114+4025047	& 166.01 & $-$0.22  & 4907.07 & 2.75 & 0.10 & 14.45 & 1.41 & 2.62\\%5
6 &2M05104317+4051399	& 166.13 & 0.68   & 4909.86 & 2.65 & $-$0.10 & 14.19 & 1.26 & 2.37\\%6
7 &2M05072939+4019254	& 166.19 & $-$0.13  & 4910.46 & 2.66 & $-$0.13 & 14.21 & 1.28 & 2.43\\%7
8 &2M05014034+4045148	& 165.18 & $-$0.75  & 4926.30 & 2.54 & $-$0.28 & 15.11 & 1.30	 & 2.60\\%8
9 &2M05113657+4102440	& 166.08 & 0.93   & 4935.57 & 2.70 & $-$0.42 & 15.42 & 1.09	 & 2.30\\%9
10 &2M05055643+4124051  & 165.16 & 0.29   & 4943.94 & 2.71 & $-$0.15 & 14.36 & 1.38 & 2.51\\%10
11 &2M05081159+4135433   & 165.25 & 0.74   & 4977.03 & 2.61 & $-$0.23 & 14.48 & 1.48	 & 2.79\\%11
12 &2M05041714+4232248  & 164.06 & 0.73   & 4977.43 & 2.76 & $-$0.33 & 13.95 & 1.21 & 2.37\\%12
13 &2M05070978+4134192  & 165.16 & 0.57   & 4982.41 & 2.76 & $-$0.27 & 14.25 & 1.26 & 2.38\\%13
14 &2M05111337+4102241  & 166.04 & 0.87   & 4989.09 & 2.75 & $-$0.03 & 13.77 & 1.32	 & 2.41\\%14
15 &2M05031839+4000169  & 165.96 & $-$0.96  & 4992.78 & 2.63 & $-$0.34 & 15.82 & 1.43	 & 3.05\\%15
16 &2M04592396+4023594  & 165.19 & $-$1.30  & 4993.97 & 2.67 & $-$0.39 & 15.14 & 1.49	 & 2.95\\%16
17 &2M05000136+4219417  & 163.75 & $-$0.02  & 5007.93 & 2.93 & $-$0.01 & 14.68 & 1.47	 & 2.58\\%17
18 &2M05102193+4121433  & 165.68 & 0.98   & 5027.16 & 2.79 & $-$0.22 & 14.39 & 1.20 & 2.39\\%18
\hline
\hline
\end{tabular}
\end{center}
\end{table}

\section{Method}

We use the color-excess ratio to express the extinction law, where the
color excess $E(\lambda_1-\lambda_2)=\Cll-\Cll^0$ is the difference
between the observed color index $\Cll$ and the intrinsic color index
$\Cll^0$. We use $E(V-\lambda_x)/E(B-V)$ ($\lambda_x: B, g, r, i, J,
H, \Ks, W1, W2$) to represent the interstellar extinction law. The key
problem is to determine the intrinsic color index $(V-\lambda_x)_0$.
Here, we will introduce two methods to obtain $(V-\lambda_x)_0$.

\subsection{Intrinsic Colors}
\subsubsection{Analytic $\Teff$--intrinsic color $\Cll^0$ relation}

Wang \& Jiang (2014) suggested that the intrinsic color index could be
represented by the bluest observed color index under some
circumstances for a given $\Teff$. This means that the intrinsic color
index $\Cll^0$ can be derived from their effective temperatures
$\Teff$ by considering the bluest star at the same $\Teff$ not
affected by reddening. They determined the $\Teff$--near-IR intrinsic
color index relation by means of a quadratic fit to the bluest stars
in the $\Teff$ vs. observed color index diagram for APOGEE K-type
giants ($3500\K\leq\Teff\leq4800\K$).
Xue et al.\ (2016) further applied this method to multiple mid-IR
bands for the APOGEE G- and K-type giants
($3600\K\leq\Teff\leq5200\K$). In order to determine the blue envelope
in the $\Teff$ vs. observed color index diagram, they adopted a
mathematical definition for the blue edge. First, they chose the
median color of the bluest 5\% of stars in bins of $\Delta
\Teff=100\K$ to represent the unreddened color. Then, they used an
exponential or quadratic function to fit the bluest color. The
original idea underlying this method was developed by Ducati (2001).

%figure B-V intrinsic color
\begin{figure}[h!]
\centering
\vspace{-0.0in}
\includegraphics[angle=0,width=6.0in]{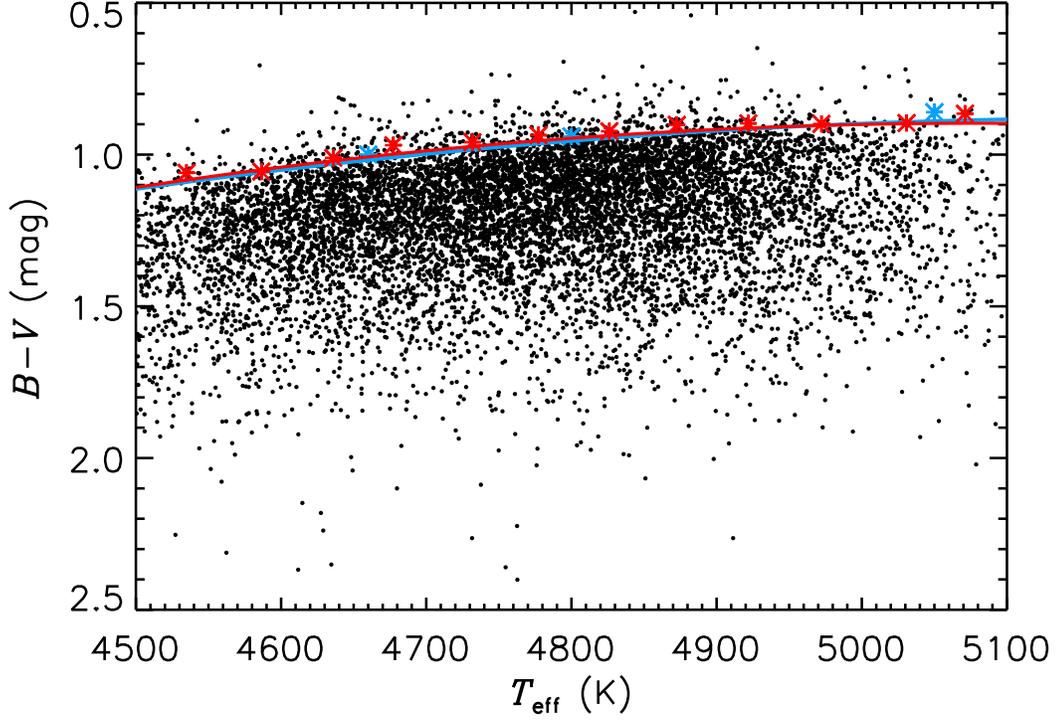}
\vspace{-0.2in}
\caption{\footnotesize
               \label{fig:BVint}
               $\Teff$ vs. observed color $(B-V)$ diagram. Black dots
               are the selected K-type giants, red asterisks show the
               median values of the bluest stars in bins of $\Delta
               \Teff=50\K$. The red solid line denotes the quadratic
               fit result to these red asterisks. For comparison, the
               blue line is the quadratic fit result to the discrete
               intrinsic color data (blue asterisks) of Johnson
               (1966).  }
\end{figure}

Since RC stars in the range $4550\K \le \Teff \le 5050\K$ are selected
as tracers for investigating the interstellar extinction, we
concentrate on the APOGEE K-type giants with $4500\K \le \Teff \le
5100\K$ to determine the intrinsic colors in the optical bands, since
the relation between IR color and $\Teff$ has already been derived
(Xue et al. 2016). The following constraints are used to obtain a
K-giant sample: (1) $4500\K \le \Teff \le 5100\K$; (2) $\log g \le
3.0$; (3) photometric uncertainties $\sigma < 0.1$ mag in the $B, V$
bands and $\sigma < 0.05$ mag in the $g, r, i, J, H, \Ks, W1, W2$
bands; (4) observed spectral signal-to-noise ratio $> 100$ and
difference between the observed and synthetic model spectra
$\chi^2_{\rm ASPCAP} < 30$; (5) [Fe/H] $> -0.5$ dex. Note that this
step employs stars from all fields, including from high Galactic
latitude areas, and is unbiased as to any specific environment. A
series of discrete median effective temperatures $\langle \Teff
\rangle$ and median observed colors $\langle \Cll \rangle$ in bins of
$\Delta \Teff=50\K$ are selected from the bluest 5\% of stars. A
quadratic function is fitted to this [$\langle \Teff \rangle$,
  $\langle \Cll \rangle$] series to determine the analytic expression
for the $\Teff$--intrinsic color relation.
Figure~\ref{fig:BVint} is the $\Teff$ vs. observed color $(B-V)$
diagram for the K-giant sample (black dots). The red asterisks are the
median values for the bluest 5\% of stars, and the red line is the
quadratic best-fitting result. For comparison, we also plot the
quadratic fit result for discrete intrinsic color data (blue
asterisks) given by Johnson (1966; blue line). The two lines are
highly consistent. The results in other bands are as follows:
\begin{equation}
\CBVint = 17.10 - 6.38 (\frac{\Teff}{10^3\K}) + 0.628 (\frac{\Teff}{10^3\K})^2 ~~,
\end{equation}
\begin{equation}
\CgVint = 11.40 - 4.31 (\frac{\Teff}{10^3\K}) + 0.424 (\frac{\Teff}{10^3\K})^2 ~~,
\end{equation}
\begin{equation}
\CVrint = 12.93 - 5.03 (\frac{\Teff}{10^3\K}) + 0.500 (\frac{\Teff}{10^3\K})^2 ~~,
\end{equation}
\begin{equation}
\CViint = 21.12 - 8.20 (\frac{\Teff}{10^3\K}) + 0.818 (\frac{\Teff}{10^3\K})^2 ~~,
\end{equation}
\begin{equation}
\CVJint = 23.57 - 8.50 (\frac{\Teff}{10^3\K}) + 0.823 (\frac{\Teff}{10^3\K})^2 ~~,
\end{equation}
where $\CBVint$, $\CgVint$, $\CVrint$, $\CViint$, and $\CVJint$
represent the intrinsic color indices $(B-V)_0, (g-V)_0, (V-r)_0$,
$(V-i)_0$, and $(V-J)_0$, respectively. Xue et al.\ (2016) already
determined the multi-band intrinsic IR colors for APOGEE G- and K-type
giants with $3600\K \le \Teff \le 5200\K$. Therefore, we adopt the IR
intrinsic colors ($J-H)_0$, $(J-\Ks)_0$, $(\Ks-W1)_0$, and
$(\Ks-W2)_0$ from their work to determine $(V-H)_0$, $(V-\Ks)_0$,
$(V-W1)_0$, and $(V-W2)_0$ (hereafter $\CVHint$, $\CVKint$,
$\CVWaint$, and $\CVWbint$, respectively).

\subsubsection{Padova stellar models}
The stellar intrinsic color can be calculated from stellar evolution
models once the metallicity, effective temperature, and surface
gravity are known. One of the most commonly used stellar evolution
models are the Padova isochrone sets of Marigo et al.\ (2008) with the
Girardi et al.\ (2010) Case A correction for low-mass, low-metallicity
AGB tracks. This isochrone set was used to determine extinction maps
toward the Milky Way bulge based on APOGEE targets by Schultheis et
al.\ (2014). We adopt very similar procedures: a step of 0.2 dex in
metallicity in the range of $-0.5 < $[Fe/H]$ < 0.5$ dex and
$\bigtriangleup (\log$ age)=0.05 [Gyr] within the range $6.6 \le \log
\rm{age/yr} \le 10.13$. Specific steps are as follows: (1) according
to the stellar [Fe/H], we derive the sequence of isochrones with the
closest, constant metallicity for each star; (2) the absolute
magnitude is derived from a two-dimensional interpolation in the
corresponding $\log g$ vs.  $\Teff$ plane, rather than based simply on
the closest data point. The interpolated value is based on a cubic
interpolation of the values of the neighboring grid points for each
$\log g$ and $\Teff$, and absolute magnitude dimension,
respectively. In this way, we derived the absolute magnitudes in ten
bands ($B, V, g, r, i, J, H, \Ks, W1, W2$) for each RC star, and the
intrinsic colors for any pair of bands are thus available.

\subsubsection{Comparison}
%

%table intrinsic color
\begin{table}[h!]
\begin{center}
\vspace{-0.3in}
\caption{\label{tab:intrinsic} Multi-band intrinsic color indices for
  the 18 RCs in the $l165^{\circ}$ region}
\vspace{0.1in}
\begin{tabular}{cccccccccccccccc}
\hline \hline
No. &$M_\Ks$ & $\CBVint$ & $\CgVint$ & $\CVrint$ & $\CViint$ & $\CVJint$
& $\CVHint$ & $\CVKint$ & $\CVWaint$ & $\CVWbint$ \\
\hline
\multicolumn{10}{c}{Analytic results}\\
\hline
1 &--& 0.992 & 0.505 & 0.337 & 0.644 & 1.81 & 2.35 & 2.45 & 2.52 & 2.43\\
2 &--& 0.932 & 0.463 & 0.297 & 0.582 & 1.71 & 2.21 & 2.30 & 2.37 & 2.29\\
3 &--& 0.925 & 0.459 & 0.293 & 0.576 & 1.70 & 2.19 & 2.28 & 2.35 & 2.27\\
4 &--& 0.922 & 0.457 & 0.291 & 0.573 & 1.70 & 2.19 & 2.27 & 2.34 & 2.26\\
5 &--& 0.916 & 0.452 & 0.287 & 0.568 & 1.68 & 2.17 & 2.25 & 2.32 & 2.24\\
6 &--& 0.915 & 0.452 & 0.287 & 0.567 & 1.68 & 2.16 & 2.25 & 2.32 & 2.24\\
7 &--& 0.915 & 0.451 & 0.287 & 0.567 & 1.68 & 2.16 & 2.25 & 2.32 & 2.24\\
8 &--& 0.912 & 0.449 & 0.285 & 0.565 & 1.68 & 2.15 & 2.24 & 2.31 & 2.23\\
9 &--& 0.910 & 0.448 & 0.284 & 0.563 & 1.67 & 2.15 & 2.23 & 2.30 & 2.22\\
10 &--& 0.908 & 0.447 & 0.283 & 0.562 & 1.67 & 2.14 & 2.22 & 2.30 & 2.22\\
11 &--& 0.903 & 0.443 & 0.281 & 0.559 & 1.66 & 2.12 & 2.20 & 2.28 & 2.20\\
12 &--& 0.903 & 0.443 & 0.281 & 0.559 & 1.66 & 2.12 & 2.20 & 2.28 & 2.20\\
13 &--& 0.903 & 0.443 & 0.281 & 0.559 & 1.66 & 2.12 & 2.20 & 2.27 & 2.20\\
14 &--& 0.902 & 0.442 & 0.280 & 0.559 & 1.65 & 2.12 & 2.20 & 2.27 & 2.19\\
15 &--& 0.901 & 0.442 & 0.280 & 0.559 & 1.65 & 2.12 & 2.19 & 2.27 & 2.19\\
16 &--& 0.901 & 0.442 & 0.280 & 0.559 & 1.65 & 2.11 & 2.19 & 2.27 & 2.19\\
17 &--& 0.900 & 0.441 & 0.280 & 0.558 & 1.65 & 2.11 & 2.19 & 2.26 & 2.18\\
18 &--& 0.898 & 0.440 & 0.279 & 0.558 & 1.64 & 2.10 & 2.18 & 2.25 & 2.18\\
\hline
\multicolumn{10}{c}{Model results}\\
\hline
1 &$-$2.17 &1.14	&0.59	&0.32	&0.58	&1.85	&2.39	&2.48	&2.52	&2.44\\
2 &$-$2.40 &1.01	&0.51	&0.26	&0.50	&1.71	&2.23	&2.30	&2.34	&2.28\\
3 &$-$1.83 &1.01	&0.52	&0.27	&0.50	&1.70	&2.20	&2.28	&2.32	&2.25\\
4 &$-$0.97 &0.95	&0.48	&0.25	&0.48	&1.68	&2.20	&2.27	&2.31	&2.26\\
5 &$-$1.74 &1.01	&0.52	&0.27	&0.49	&1.68	&2.17	&2.25	&2.28	&2.21\\
6 & --  & --	&--	&--	&--	&--	&--	&--	&--	&--\\
7 &$-$1.81 &0.98	&0.5	&0.26	&0.48	&1.67	&2.16	&2.24	&2.27	&2.21\\
8 &$-$2.18 &0.95	&0.48	&0.25	&0.48	&1.66	&2.15	&2.23	&2.26	&2.2\\
9 &$-$1.19 &0.92	&0.47	&0.24	&0.46	&1.64	&2.14	&2.21	&2.24	&2.2\\
10 &-- & --	&--	&--	&--	&--	&--	&--	&--	&--\\
11 &$-$2.00 &0.94	&0.47	&0.24	&0.46	&1.62	&2.10	&2.17	&2.2	&2.15\\
12 &$-$1.34 &0.92	&0.46	&0.24	&0.46	&1.62	&2.11	&2.18	&2.21	&2.16\\
13 &$-$1.41 &0.93	&0.47	&0.24	&0.46	&1.62	&2.10	&2.17	&2.20	&2.15\\
14 &$-$1.74 &0.96	&0.49	&0.25	&0.46	&1.61	&2.09	&2.16	&2.19	&2.13\\
15 &$-$1.80 &0.92	&0.46	&0.23	&0.45	&1.60	&2.08	&2.15	&2.18	&2.14\\
16 &$-$1.65 &0.91	&0.46	&0.23	&0.45	&1.60	&2.09	&2.15	&2.18	&2.14\\
17 &$-$1.11 &0.96	&0.49	&0.24	&0.45	&1.61	&2.08	&2.15	&2.19	&2.13\\
18 &$-$1.48 &0.91	&0.46	&0.24	&0.45	&1.59	&2.06	&2.13	&2.16	&2.11\\

\hline \hline
\end{tabular}
\end{center}
\end{table}

%figure intrinsic color
\begin{figure}[h!]
\centering
\vspace{-1.4in}
%\hspace{-0.5in}
\includegraphics[angle=90,width=6.5in]{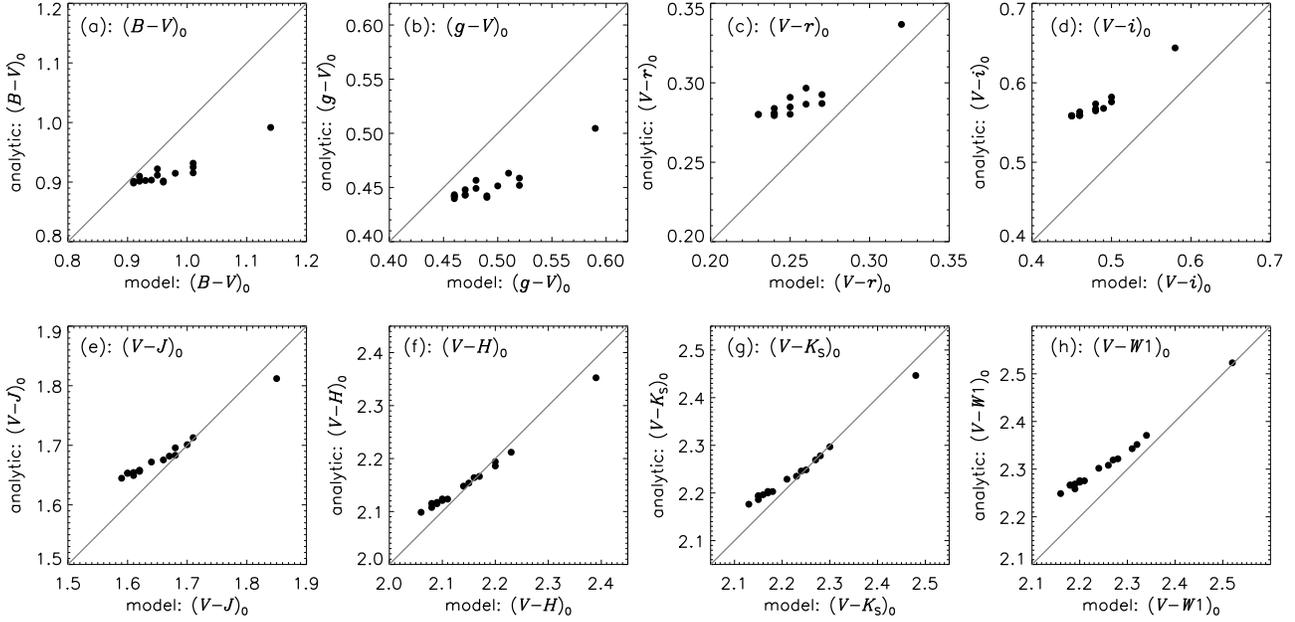}
\vspace{-1.1in}
\caption{\footnotesize
               \label{fig:int}
           Comparison of the 18 RC intrinsic color indices derived
           from the analytic $\Teff$--$C_{\lambda1\lambda2}^0$
           relation with those derived from the Padova isochrone
           models. The values are listed in Table~\ref{tab:intrinsic}.
}
\end{figure}

The intrinsic color indices of the 18 RC stars in the $l165^{\circ}$
region are included in Table~\ref{tab:intrinsic}. The top part
displays the intrinsic values derived from the analytic
$\Teff$--$C_{\lambda1\lambda2}^0$ relation; the bottom part displays
the values derived from the Padova stellar models. The $\Ks$-band
absolute magnitudes from the Padova stellar models are also listed in
the first column of Table~\ref{tab:intrinsic}. They range from $\Ks =
-0.97 \magni$ to $-2.40 \magni$. The average $\Ks$-band absolute
magnitude value, $\Ks = -1.68 \magni$, is slighter smaller than the
typical RC value, $-1.54 \magni$ -- $-1.61 \magni$, which results in a
higher extinction value. We also compare the intrinsic color indices
determined based on these two methods. Figure~\ref{fig:int} shows the
comparison. The vertical axis shows the intrinsic color indices
derived from the analytic $\Teff$--$C_{\lambda1\lambda2}^0$ relation,
and the horizontal axis shows the values derived from the Padova
models. The intrinsic IR color indices, i.e., $\CVJint$, $\CVHint$,
$\CVKint$, and $\CVWaint$, are internally consistent. However,
Figure~\ref{fig:int} shows that the optical color indices exhibit
notable differences: for $\CBVint$ and $\CgVint$, the analytic results
are lower than the model results; for $\CVrint$ and $\CViint$, the
analytic results are higher than the model results. It is unclear
whether these differences are caused by flaws in the stellar models or
by the analytic method. Nevertheless, the difference is mostly smaller
than 0.05 $\magni$ in color, comparable to the photometric
uncertainties.
\subsection{Color-Excess Ratio}

With the observed color index (the difference between two observed
magnitudes) and the intrinsic color index (derived from its dependence
on $\Teff$ or on isochrone sets), the color excess can be calculated
easily. The color excesses $E(V-\lambda_x)$ ($\lambda_x: B, g, r, i,
J, H, \Ks, W1, W2$) were derived for each sample star. In principle,
the color-excess ratio, e.g., $E(V-\lambda_x)/E(B-V)$, can be regarded
as indicator of the extinction law.

\section{Results and Discussion}
\subsection{Optical--Mid-IR Extinction}\label{ext}
%

%table color excess ratio
\begin{table}[h!]
{\footnotesize
\begin{center}
\vspace{-0.3in}
\caption{\label{tab:ratio} Multi-band color-excess ratios for the 18
  RC stars in the $l165^{\circ}$ region}
\vspace{0.1in}
\begin{tabular}{cccccccccccccccc}
\hline \hline
No. & $\EBV$ & $\EgV/\EBV$  & $\EVr/\EBV$ & $\EVi/\EBV$ & $\EVJ/\EBV$
& $\EVH/\EBV$  & $\EVK/\EBV$ & $\EVWa/\EBV$ & $\EVWb/\EBV$\\
\hline
\multicolumn{10}{c}{Analytic results}\\
\hline
1 & 0.448 &0.398	& 0.731 & 1.22	& 2.13	& 2.45	& 2.62	& 2.73	& 2.72\\
2 & 0.598 &0.599	& 0.455 & 0.980	& 1.87	& 2.12	& 2.33	& 2.44	& 2.45\\
3 & 0.540 &0.501	& 0.314 & 0.690	& 1.31	& 1.59	& 1.63	& 1.79	& 1.78\\
4 & 0.319 &0.459	& 0.428 & 0.952	& 1.55	& 1.76	& 2.02	& 2.18	& 2.25\\
5 & 0.493 &0.441	& 0.563 & 1.04	& 1.91	& 2.13	& 2.31	& 2.44	& 2.44\\
6 & 0.347 &0.387	& 0.692 & 1.12	& 1.97	& 2.28	& 2.47	& 2.66	& 2.67\\
7 & 0.362 &0.481	& 0.605 & 1.03	& 2.07	& 2.39	& 2.66	& 2.64	& 2.73\\
8 & 0.386 &0.583	& 0.701 & 1.33	& 2.40	& 2.78	& 3.03	& 3.21	& 3.27\\
9 & 0.184 &0.448	& 1.309 & 2.17	& 3.43	& 4.16	& 4.54	& 4.72	& 4.78\\
10 &0.476 &0.610	& 0.433 & 0.942	& 1.76	& 2.03	& 2.22	& 2.29	& 2.28\\
11 &0.575 &0.554	& 0.472 & 0.991	& 1.96	& 2.28	& 2.48	& 2.59	& 2.62\\
12 &0.308 &0.482	& 0.716 & 1.30	& 2.33	& 2.62	& 2.79	& 3.05	& 3.07\\
13 &0.361 &0.596	& 0.501 & 1.01	& 1.99	& 2.32	& 2.58	& 2.69	& 2.67\\
14 &0.416 &0.636	& 0.489 & 0.992	& 1.81	& 2.11	& 2.23	& 2.35	& 2.36\\
15 &0.533 &0.561	& 0.650 & 1.31	& 2.62	& 3.03	& 3.31	& 3.41	& 3.49\\
16 &0.586 &0.591	& 0.512 & 1.08	& 2.21	& 2.59	& 2.75	& 2.89	& 2.93\\
17 &0.573 &0.572	& 0.384 & 0.828	& 1.62	& 1.90	& 2.06	& 2.15	& 2.17\\
18 &0.303 &0.544	& 0.698 & 1.40	& 2.47	& 2.90	& 3.07	& 3.20	& 3.25\\
\hline
\multicolumn{10}{c}{Model results}\\
\hline
1	&0.300 &0.309	&1.15	&2.04	&3.05	&3.54	&3.81	&4.09	&4.03\\
2	&0.520 &0.599	&0.594	&1.29	&2.16	&2.41	&2.68	&2.86	&2.83\\
3	&0.455 &0.460	&0.423	&0.99	&1.56	&1.88	&1.93	&2.19	&2.15\\
4	&0.291 &0.423	&0.609	&1.36	&1.75	&1.88	&2.21	&2.51	&2.47\\
5	&0.399 &0.374	&0.739	&1.48	&2.37	&2.63	&2.85	&3.12	&3.10\\
6    & --	& --	&--	&--	&--	&--	&--	&--	&--\\
7	&0.297 &0.423	&0.828	&1.56	&2.56	&2.93	&3.26	&3.38	&3.42\\
8	&0.348 &0.559	&0.878	&1.72	&2.71	&3.10	&3.38	&3.70	&3.71\\
9	&0.174 &0.348	&1.64	&2.90	&3.81	&4.45	&4.92	&5.36	&5.20\\
10  & --	& --	&--	&--	&--	&--	&--	&--	&--\\
11	&0.538 &0.542	&0.580	&1.24	&2.17	&2.48	&2.71	&2.91	&2.89\\
12	&0.291 &0.452	&0.897	&1.72	&2.59	&2.82	&3.03	&3.45	&3.38\\
13	&0.334 &0.563	&0.664	&1.38	&2.26	&2.57	&2.88	&3.13	&3.02\\
14	&0.358 &0.606	&0.653	&1.43	&2.23	&2.53	&2.70	&2.96	&2.92\\
15	&0.514 &0.546	&0.771	&1.57	&2.82	&3.21	&3.52	&3.70	&3.72\\
16	&0.577 &0.569	&0.606	&1.28	&2.34	&2.67	&2.87	&3.08	&3.07\\
17	&0.513 &0.543	&0.506	&1.14	&1.88	&2.18	&2.37	&2.54	&2.53\\
18	&0.291 &0.496	&0.861	&1.83	&2.75	&3.15	&3.35	&3.64	&3.60\\
\hline \hline
\end{tabular}
\end{center}
}
\end{table}
%

%figure extinction_color excess ratio
\begin{figure}%[h!]
\vspace{-0.8in}
\hspace{-0.6in}
\begin{minipage}[t]{0.5\linewidth}
\centering
\includegraphics[angle=0,width=10.3in]{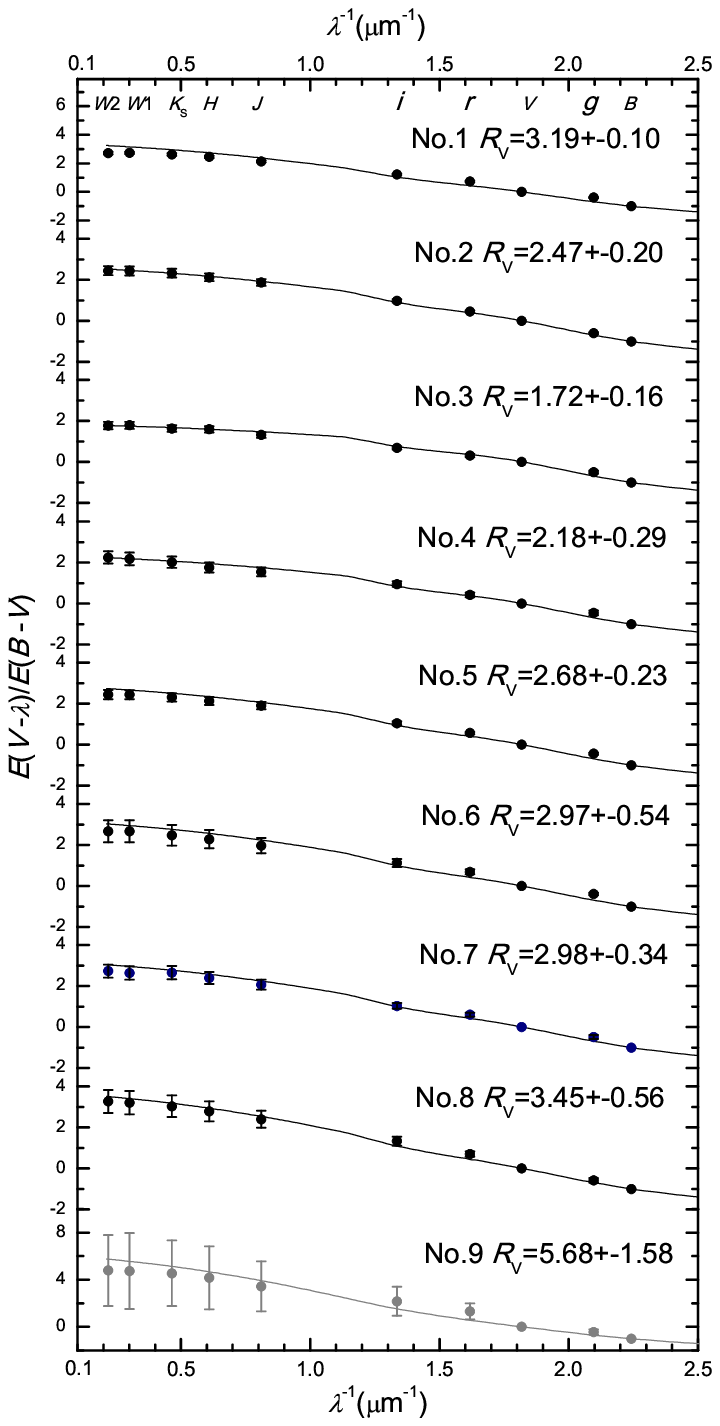}
\end{minipage}
\begin{minipage}[t]{0.5\linewidth}
\centering
\includegraphics[angle=0,width=10.3in]{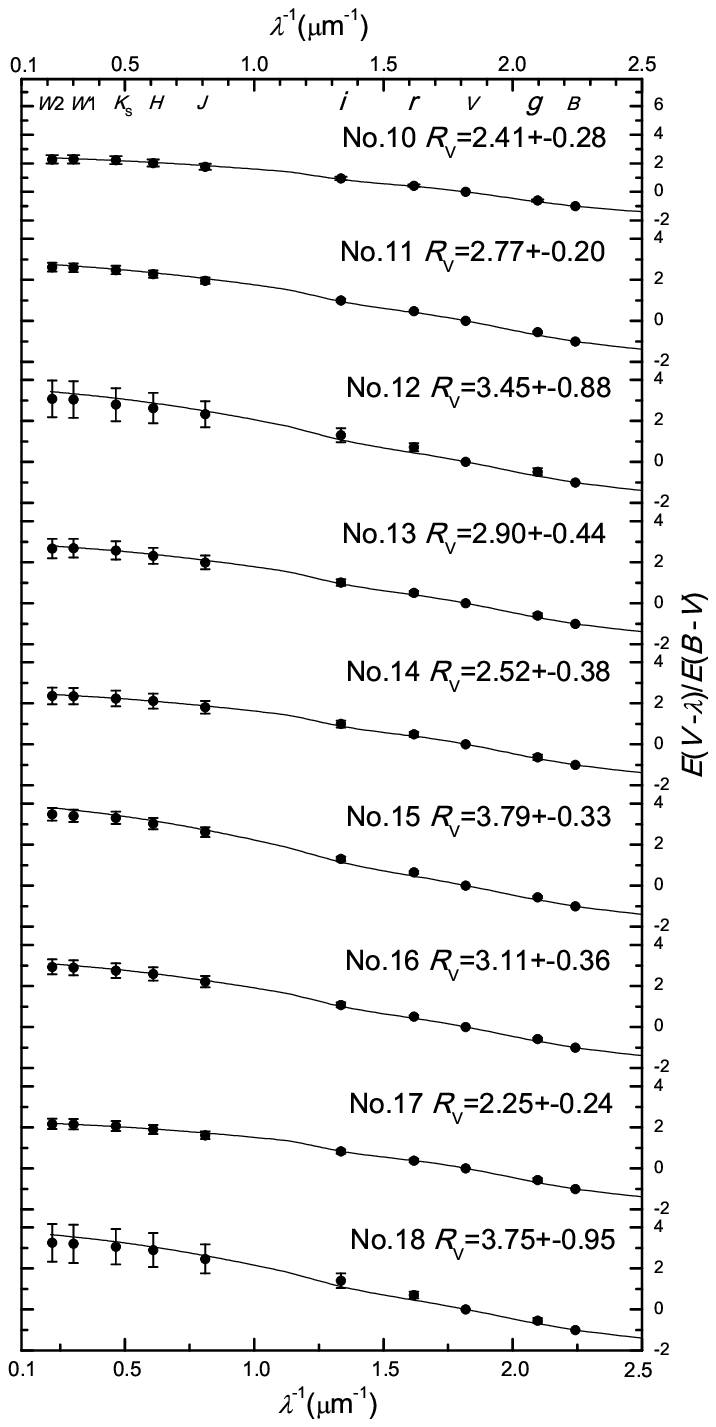}
\end{minipage}
\vspace{-12.0in}
\caption{\footnotesize
               \label{fig:ratio}
           Multi-band extinction and corresponding $\Rv$ values for
           the 18 RC stars. Color excesses have been determined by
           adopting intrinsic colors based on the analytic
           $\Teff$--$C_{\lambda1\lambda2}^0$ relation. The black line
           shows the best-fitting results based on using CCM89's
           equations. The error bar associated with the $\Rv$ values
           were derived based on 20,000 Monte Carlo simulations.  }
\end{figure}

Using the method described above, the color-excess ratios
$E(V-\lambda_x)/E(B-V)$ were derived for the 18 RC stars in the
$l165^{\circ}$ region. The results are tabulated in
Table~\ref{tab:ratio}. Figure~\ref{fig:ratio} displays the variations
in the color-excess ratios with waveband, where the color-excess
ratios were derived from the analytic
$\Teff$--$C_{\lambda1\lambda2}^0$ relation.

%****CCM89****fitting to get Rv******
The derived extinction law is fitted with the CCM89 equations. Since
CCM89 suggested that the extinction law can be described by $\Rv$,
this yields the best-fitting $\Rv$ value for the prevailing extinction
law. In practice, steps of 0.001 are adopted for $0.5 \le \Rv \le
7.0$. For each $\Rv$, $A_{\lambda_x}/\AV$ is calculated based on the
CCM89 equations in all ten bands, from which the color-excess ratio
$E(V-\lambda_x)/E(B-V)$ is derived. The best-fitting $\Rv$ is
determined by assessment of the minimum chi-squared value between the
color-excess ratio derived using the CCM89 equations and this work.

The $\Rv$ values and best-fitting CCM89 lines are shown in
Figure~\ref{fig:ratio}. The CCM89 extinction curve can fit all
cases. Table~\ref{tab:ratio} shows that the $\EBV$ and $\EgV$ of RC
star No.\ 9 are rather small and the color-excess ratios $\EVl/\EBV$
in the $r, i, J, H, \Ks, W1$, and $W2$ bands are all apparently larger
than the values for the other 17 RC stars. This is caused by the
extremely small $\EBV$ value, which may be owing to unreliable
calibrations in the $B, V$, and $g$ bands. This star was removed from
further analysis.

%the variation of extinction
The mean $\Rv$ value of the remaining 17 RCs is 2.8, which is close to
$\Rv = 3.1$, commonly adopted for the average extinction law toward
Galactic diffuse clouds. 
Schlafly \& Finkbeiner (2011) measured reddening values 
for the diffuse ISM based on a large 
sample of SDSS sources, and found an average extinction law 
consistent with $\Rv=3.1$.
On the other hand, as can be seen from
Figure~\ref{fig:ratio}, there is a clear diversity of $\Rv$ values
among stars even in such a small region, which was carefully selected
to be very diffuse along any of its sightlines. The lowest $\Rv$ value
is $\Rv=1.72$ (No.\ 3); the highest value reaches $\Rv=3.79$
(No.\ 15). This diversity significantly exceeds the intrinsic errors
(see the next section), and so it is likely real. Moreover, 
the lowest value of $\Rv=1.72$ is smaller than the previously published 
lowest value of $\Rv=2.1$ towards the HD 210101 sightline. 
The reddening towards Type Ia supernovae 
commonly shows a low $\Rv$ value $<$ 2.0 (Howell 2011, and references therein), 
in some cases smaller than 1.0, which means the possibility for 
$\Rv$ being smaller than 1.7. If we search in larger sample sizes, 
even lower $\Rv$ values may be found in the Milky Way.
As for the IR bands, there is little variation in $E(J-H)/E(J-\Ks)$, 
with a mean around 0.64, which agrees with the result of Wang \& Jiang (2014). 
The $W1$ and $W2$ bands seem to show some diversity, 
but with less confidence because of their low sensitivity.

\subsection{Error Analysis}

The error in the color-excess ratios $[E(V-\lambda_x)/E(B-V)]_{\rm
  err}$ originates from a few contributors, including the observed and
intrinsic color indices. By constraining the photometric quality of
our sample stars to $\sigma < 0.1$ mag in the $B, V$ bands and $\sigma
< 0.05$ mag in the $g, r, i, J, H, \Ks, W1$, and $W2$ bands, the
average photometric error is $\sim$ 0.06 mag in $B$, $\sim$ 0.04 mag
in $V$, and $\sim$ 0.02 mag in $g, r, i, J, H, \Ks, W1$, and $W2$
\footnote{For the 18 RC tracers, 
the APASS/BV magnitudes and the XSTPS-GAC/gr magnitudes are compared with 
the Pan-STARRS1(PS1, Hodapp et al.\ 2004)/gr magnitudes, 
the most similar filters in the system. 
We found that APASS/B $\sim$ PS1/g $+$0.78 mag with a standard deviation of 0.07 mag, APASS/V $\sim$ PS1/r $+$0.40 mag with a standard deviation of 0.07 mag,  
XSTPS-GAC/g $\sim$ PS1/g $+$0.12 mag with a standard deviation of 0.03 mag, 
and XSTPS-GAC/r $\sim$ PS1/r $-$0.10 mag with a standard deviation of 0.03 mag. 
The systematical shift is caused by the difference of the filters. 
The standard deviation results from the photometric error of the two programs. 
For the APASS photometry, a sum of 0.07 mag is consistent with our statistical uncertainty of 0.06 mag in B band and 0.04 mag in V band. 
For the XSTPS-GAC photometry, the 0.03 mag deviation is smaller than 0.05 mag 
that is our data quality requirement in the g and r band.  
Meanwhile, the largest deviation occurs for RC star No. 9. 
In Section 5.1, we argued that this star may have an unreliable calibration 
in the B, V, and g bands, and we removed it from further analysis. 
Therefore, the APASS and the XSTPS-GAC data is reliable for the other 17 sources.
}.
Consequently, the average error in the observed color index is $\sim$
0.1 mag for $\CBV$ and $\sim$ 0.06 mag for $\CgV$, $\CVr$, $\CVi$,
$\CVJ$, $\CVH$, $\CVK$, $\CKWa$, and $\CKWb$.
The average error in the APOGEE $\Teff$ is $\sim50 \K$. Using the
$\Teff$---intrinsic color index relation to derive the intrinsic
colors from Eqs (1)--(6), the $\Teff$ error causes an average error of
$\sim$ 0.006 mag for $\CBVint$, $\sim$ 0.004 mag for $\CgVint$, $\sim$
0.003 mag for $\CVrint$, $\sim$ 0.004 mag for $\CViint$, $\sim$ 0.013
mag for $\CVJint$, $\sim$ 0.022 mag for $\CVHint$, $\sim$ 0.023 mag
for $\CVKint$, $\sim$ 0.023 mag for $\CVWaint$, and $\sim$ 0.025 mag
for $\CVWbint$.
Combining the photometric and intrinsic color errors, the
uncertainties in the color excesses are $(\EBV)_{\rm err} \sim 0.11$
mag, $(\EgV)_{\rm err} \sim 0.064$ mag, $(\EVr)_{\rm err} \sim 0.063$
mag, $(\EVi)_{\rm err} \sim 0.064$ mag, $(\EVJ)_{\rm err} \sim 0.073$
mag, $(\EVH)_{\rm err} \sim 0.082$ mag, $(\EVK)_{\rm err} \sim 0.083$
mag, $(\EVWa)_{\rm err} \sim 0.083$ mag, and $(\EVWb)_{\rm err} \sim
0.085$ mag.

Having thus determined the errors in the color excesses, we use Monte
Carlo simulations to calculate the statistical uncertainties in the
color-excess ratios $[E(V-\lambda_x)/E(B-V)]_{\rm err}$ and in the
total-to-selective ratio $(\RV)_{\rm err}$. Taking the error in the
color excess into account, we performed 20,000 simulations. A Gaussian
function was used to fit the distributions of $E(V-\lambda_x)/E(B-V)$
and $\RV$. The widths of the Gaussian distributions were considered to
represent the errors in the color-excess ratios
$[E(V-\lambda_x)/E(B-V)]_{\rm err}$ and the error in the reddening
$(\RV)_{\rm err}$.

\begin{figure}%[h!]
\vspace{-0.0in}
\hspace{-0.4in}
\begin{minipage}[t]{0.52\linewidth}
\centering
\includegraphics[angle=0,width=10.0in]{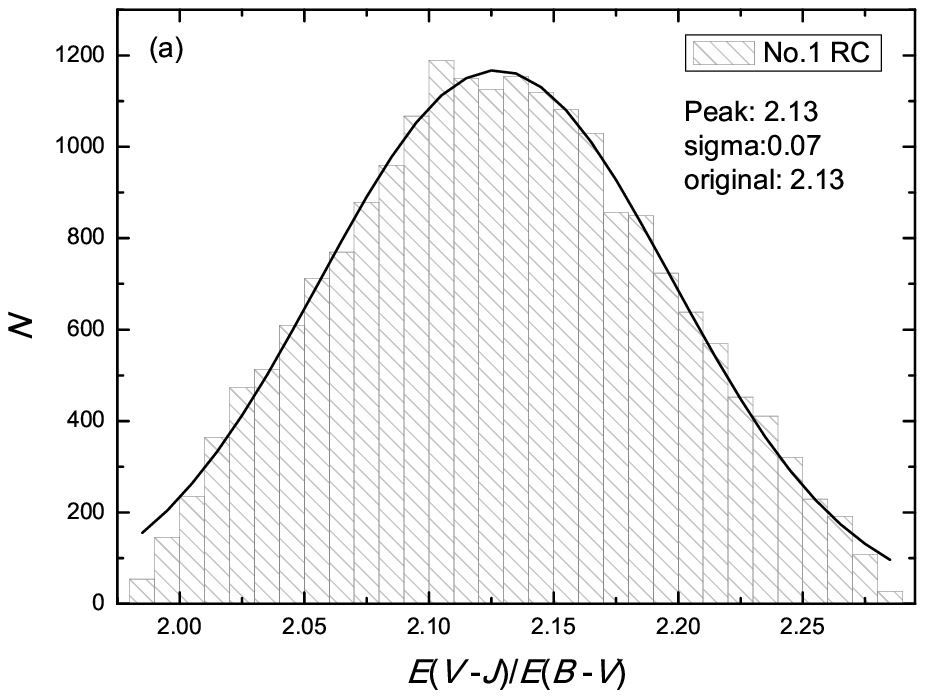}
\end{minipage}
\begin{minipage}[t]{0.52\linewidth}
\centering
\includegraphics[angle=0,width=10.0in]{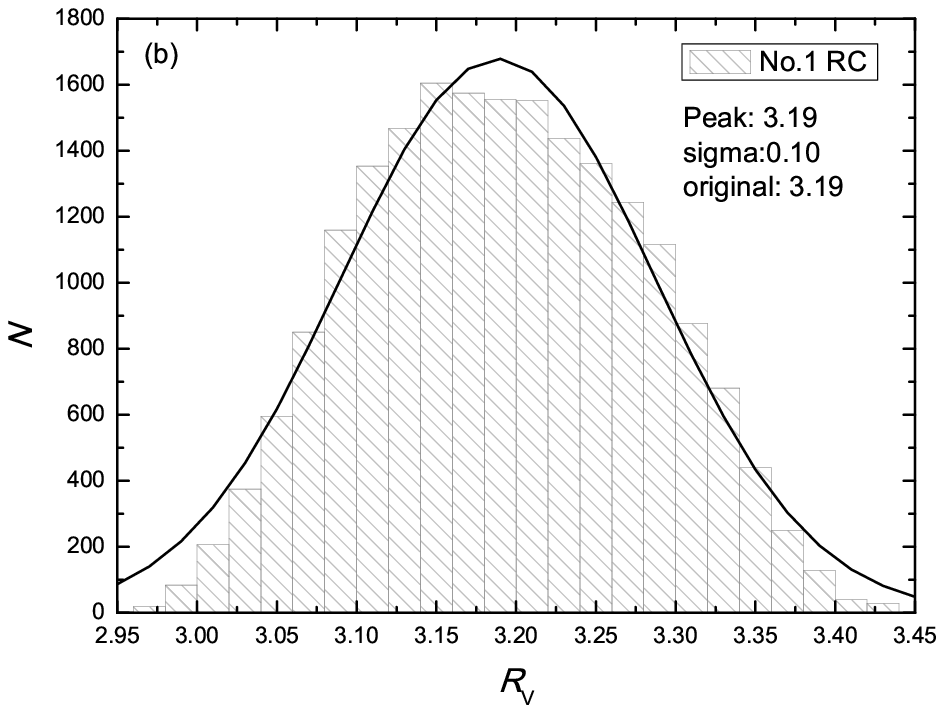}
\end{minipage}
\vspace{-4.3in}
\caption{\footnotesize
               \label{fig:simu}
          (a) Color-excess ratio $\EVJ/\EBV$ distribution and (b)
               reddening $\RV$ distributions of 20,000 Monte Carlo
               resampling results for RC star No.\ 1. }
\end{figure}

The distributions of $E(V-J)/E(B-V)$ and $\RV$ resulting from the
Monte Carlo resampling for RC star No.\ 1 are shown in
Figure~\ref{fig:simu} as an example. The distributions of
$E(V-J)/E(B-V)$ and $\RV$ are both well-fitted by a Gaussian function,
with a peak at 2.13 and a width of 0.07, and with a peak at 3.19 and a
width of 0.10, respectively. In comparison, our best-fitting results
$E(V-J)/E(B-V)=2.13$ and $\RV=3.20$ are highly consistent with the
Monte Carlo simulation results. The Monte Carlo method reconfirms our
fits. The results of our error analysis for the 18 RC stars are
presented in Table~\ref{tab:error}. Except for RC star No.\ 9, the
average error in $\RV$ is about 13.4\%. In comparison, the range of
the derived $\Rv$ values, from 1.7 to 3.8, is much larger than the
typical error, and thus the variation in $\Rv$ is real and cannot be
attributed merely to errors.

%figure for simulation

\begin{sidewaystable}[h!]
\begin{center}
\footnotesize
\caption{\label{tab:error} 
  Results of our error analysis for the 18 RC stars in the $l165^{\circ}$ region}
\vspace{0.0in}
\begin{tabular}{lcccccccccccccccccc}
\hline \hline
No.  & 1 & 2 & 3 & 4 & 5 & 6
& 7 & 8 & 9 & 10 & 11 & 12
& 13  & 14 & 15 & 16 & 17 & 18\\
\hline
$\RV$ &3.19	&2.47	&1.72	&2.18	&2.68	&2.97	
           &2.98	&3.45	&5.68	&2.41	&2.77	&3.45	
           &2.90	&2.52	&3.79	&3.11	&2.25	&3.75\\
$(\RV)_{\rm err}$ &0.10	&0.20	&0.16	&0.29	&0.23	&0.54	
                            &0.34	&0.56	&1.58	&0.28	&0.20	&0.88	
                            &0.44	&0.38	&0.33	&0.36	&0.24	&0.95\\
Peak($\RV$) &3.19	&2.47	&1.74  &2.17	&2.68	&2.90	
                      &2.95	&3.41	&5.10	&2.31	&2.68	&3.22	
                      &2.71	&2.37	&3.73	&2.99	&2.15	&3.36\\
\hline
\hline
\end{tabular}
\end{center}
\end{sidewaystable}

\subsection{Distances to the RC stars}

Variations in $\Rv$ are usually related to the interstellar
environment, whether diffuse or dense. Although the sightlines in the
$l165^{\circ}$ region are essentially diffuse, the extinction law
still exhibits significant variations. A possible reason may be that
the degree of diffusivity varies along different sightlines. In order
to quantify the interstellar environment, we derived the distances to
the sample stars so that the specific extinction per kiloparsec (kpc)
can be measured.
%

%figure for distance
\begin{figure}%[h!]
\vspace{-0.0in}
\hspace{-0.5in}
\begin{minipage}[t]{0.51\linewidth}
\centering
\includegraphics[angle=0,width=3.8in]{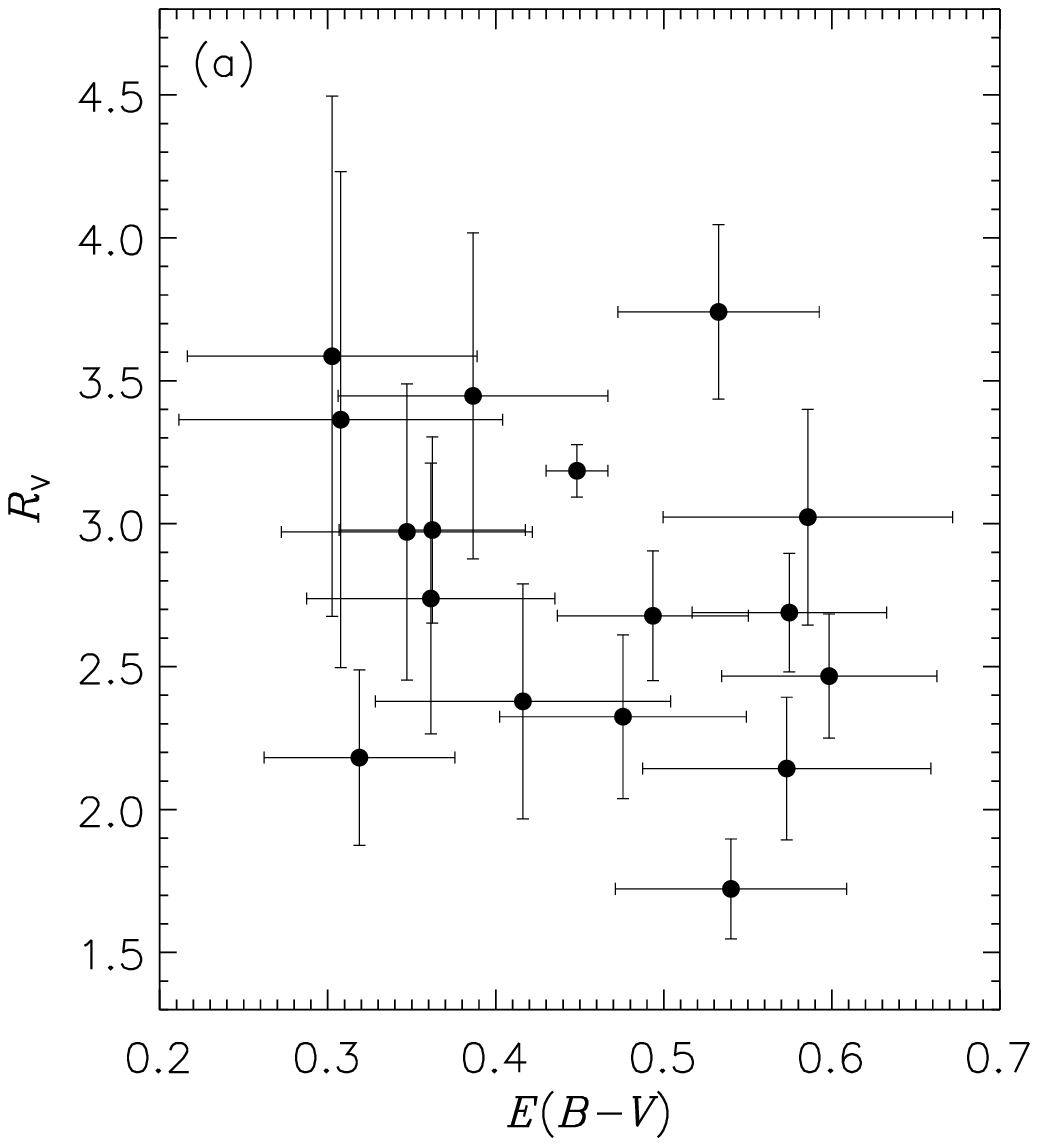}
\end{minipage}
\begin{minipage}[t]{0.51\linewidth}
\centering
\includegraphics[angle=0,width=3.8in]{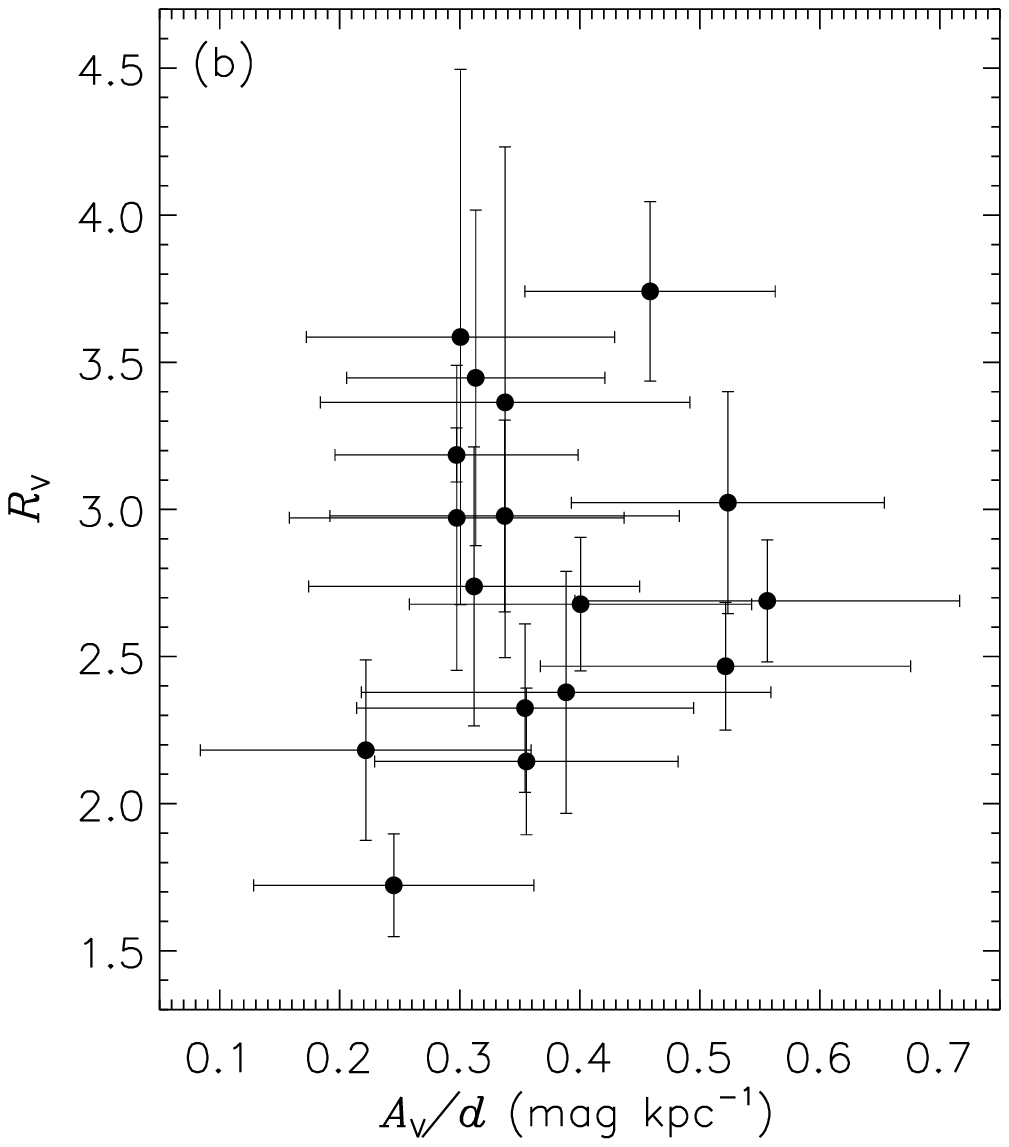}
\end{minipage}
\vspace{-0.0in}
\caption{\footnotesize
               \label{fig:dist}
          Distribution of reddening $\RV$ with (a) color excess $\EBV$
          and (b) specific visual extinction per kiloparsec $\AV/d$
          for the 17 RC stars in the $l165^{\circ}$ region.  }
\end{figure}

RC stars are good standard candles for estimating astronomical
distances, since their absolute luminosities are fairly independent of
stellar composition or age. In particular, the near-IR $I$- and
$\Ks$-band luminosities have been widely used to retrieve RC
distances. Alves (2000) was the first to consider using the RC
$\Ks$-band magnitude as a distance indicator. He found $M_{\Ks}=
-1.61\pm0.03\magni$, with a weak dependence on metallicity.
Groenewegen (2008) reinvestigated the absolute magnitude of the RC
stars based on revised {\sl Hipparcos} parallaxes. He obtained
$M_{\Ks}= -1.54\pm0.04\magni$, which is slightly fainter than
previously published values. In addition, Nishiyama et al.\ (2006)
used $M_{\Ks}= -1.59\magni$ based on Bonatto et al.\ (2004) to measure
the distance to the Galactic Center. The distance to the Galactic
Center measured this way is in agreement with results obtained based
on other methods (e.g., de Grijs \& Bono 2016). Therefore, we adopt
$M_{\Ks}= -1.59\magni$ to estimate the distances to our RC stars.

In order to calculate the distance $d$ using $\log d = 1 + 0.2(m_\Ks -
M_\Ks - A_\Ks)$, we need to know $A_\Ks$, the $\Ks$-band extinction.
As we have derived a series of color excesses $E(V-\lambda_x)$ and the
relative extinction $\Alx/\AV$ using CCM89's equations, a series of
corresponding $\AV$ values can be obtained: $\AV =
E(V-\lambda_x)/(1-\Alx/\AV)$. The median $\langle \AV \rangle$ of this
series is taken as the absolute $V$-band extinction. Then,
$\AKs=\AKs/\AV \times \langle \AV \rangle$ is obtained. The derived
distances range from 2.67 kpc to 4.49 kpc. The average $\AV$ per
kiloparsec is 0.37 mag kpc$^{-1}$, based on taking all 17 RC stars in
this region. Considering that the average rate of interstellar
extinction in the $V$ band is usually taken as 0.7--1.0 mag
kpc$^{-1}$, simply based on observations in the solar neighborhood
(Gottlieb \& Upson 1969; Milne \& Aller 1980), a gradient of 0.37 mag
kpc$^{-1}$ means that we are dealing with a really diffuse medium. The
highest specific extinction in our region of interest is approximately
0.56 mag kpc$^{-1}$, still significantly below the average rate.

Since the color excess $E(B-V)$ represents in general the density of dust, 
larger $\Rv$ is expected at larger $E(B-V)$, 
like dense clouds usually show high $\Rv$ values.
We investigate the variation in $\RV$ with $E(B-V)$ shown in Figure~\ref{fig:dist}a. 
The Pearson correlation coefficient is $-0.34$, indicating no correlation between $\Rv$ and $E(B-V)$ in the $E(B-V)$ range from 0.2 to 0.6 mag. 
Schlafly et al.\ (2016) measured optical--IR extinction curve spatial variations 
to tens of thousands of APOGEE stars. They also found that the variation in $\Rv$ 
is uncorrelated with dust column density up to $E(B-V) \approx 2$ mag.
This non-correlation seems to contradict with normal expectation. 
However, dense molecular clouds exhibit high dust extinction 
with $\AV >$ 5--10 mag (Snow \& McCall 2006), 
corresponding to $E(B-V)$ $> 1.6$ mag (for $\RV=3.1$). 
Therefore, these RC sightlines are still diffuse regions. 
Dense regions are needed to investigate a possible correlation 
between $\Rv$ and $E(B-V)$.  
On the other hand, a large E(B-V) may result from a pile of diffuse clouds. 
The specific extinction per distance would be a better measure of the ISM environment. 
Figure~\ref{fig:dist}b displays the variation in $\RV$ with
specific visual extinction per kpc, $\AV/d$. The error bars were determined 
using error-propagation theory and based on Monte Carlo simulations. 
Since the Pearson correlation coefficient between $\RV$ and $\AV/d$ is 0.14, 
There is no apparent trend for $\RV$ with $\AV/d$.  
If we were to adopt different $M_{\Ks}$ values, $M_{\Ks}$ = $-1.54\magni$
or $-1.61\magni$, the distance would change by about 3\%, 
but this change is systematic and would
therefore not change our conclusion, i.e., that the variation in $\RV$
is independent of both $E(B-V)$ and $\AV/d$ within this small range of
$E(B-V)(< 0.6)$ mag.

\subsection{Effects of Metallicity}
%

%figure metallicity
\begin{figure}[h!]
\centering
\vspace{-1.4in}
\includegraphics[angle=0,width=6.0in]{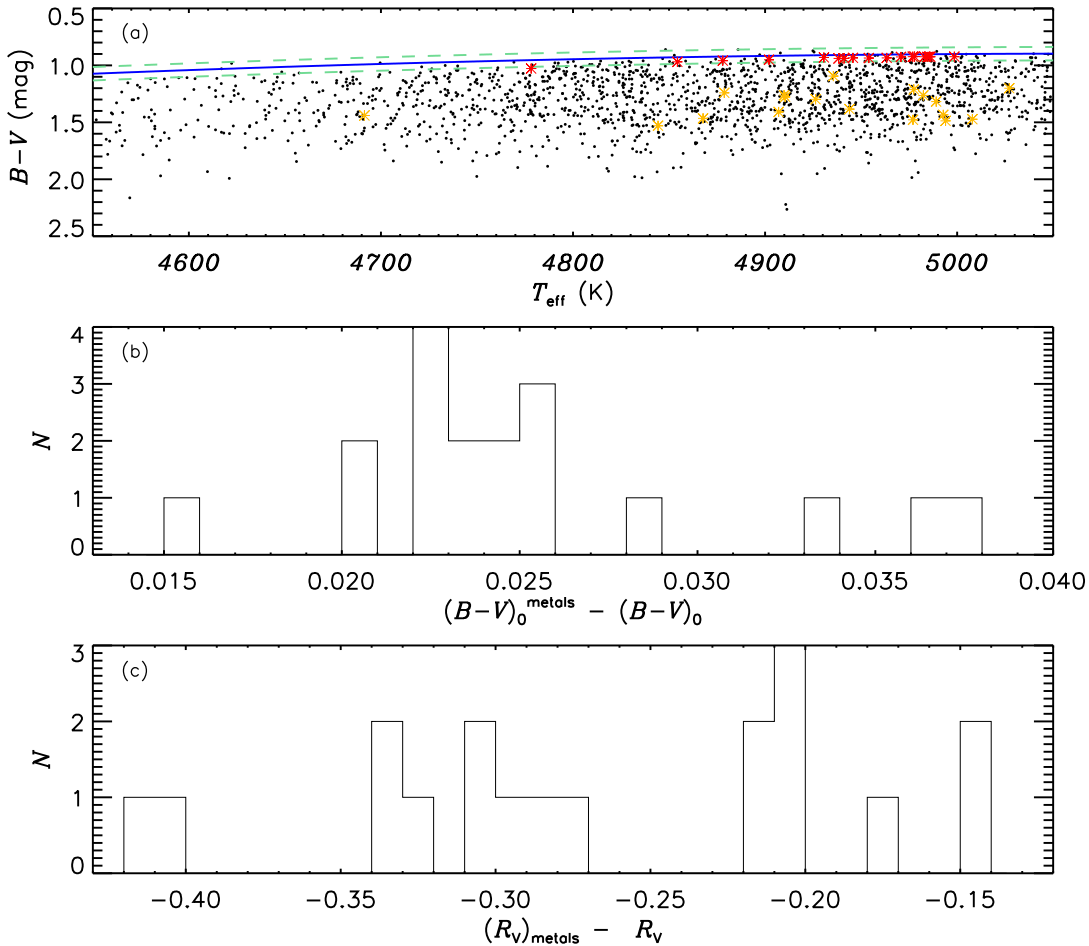}
\vspace{0.4in}
\caption{\footnotesize
               \label{fig:metallicity}
           Effects of metallicity on intrinsic RC colors. (a) Similar
           stars to the 18 target RC stars in the effective
           temperature $\Teff$ vs. observed color index ($B-V$)
           diagram. The red asterisks are the average values for stars
           that are similar to each of the 18 target RC stars (yellow
           asterisks) in $l165^\circ$; the blue solid line denotes the
           $\Teff$--$\CBVint$ relation, with the green dashed line
           representing the 1$\sigma$ envelope. (b) Comparison of the
           intrinsic colors derived above with those determined from
           the $\Teff$--$\CBVint$ intrinsic color relation. (c)
           Comparison of the $\Rv$ values determined from the
           intrinsic color indices derived above and those derived
           from the $\Teff$--intrinsic color relation.}
\end{figure}

%effect of metallicity
The extinction toward the $l165^{\circ}$ region is relatively small,
with $\AV$ ranging from about 0.7 to 2.0 mag, i.e., $E(B-V)$ ranges
from about 0.2 to 0.6 mag. For such a low extinction, a small
variation in color excess would lead to a significant difference in
the color-excess ratio $E(V-\lambda_x)/E(B-V)$. Metallicity is a
secondary parameter affecting the intrinsic color index, in addition
to the effective temperature. In our derivation of the intrinsic
colors (Section 4.1.1), metallicity was not taken into
account. Therefore, here we discuss the effect of metallicity on the
intrinsic colors of RC stars. For each of the 18 target RC stars,
stars were selected that have similar stellar parameters (including
metallicities) to the target stars and are located `close to' the
analytic best-fitting $\Teff$--$\CBVint$ intrinsic color line.
Instead of reading the intrinsic color index from the best-fitting
line, the average values of the observed color indices of these
similar stars were taken as the intrinsic color indices for each
target RC star. The `similar' stellar parameters are defined as those
parameters that are found within the ranges $\Teff(\rm
target)\pm100\K$, $\log g(\rm target)\pm0.2$, and $[Fe/H](\rm
target)\pm0.4$ dex.\footnote{Based on the uncertainties in stellar
  parameters, the parameter ranges of the target stars should be
  $\Teff(\rm target)\pm50\K$, $\log g(\rm target)\pm0.2$, and
  $[Fe/H](\rm target)\pm0.2$. In order to ensure that each of the 18
  target stars has at least one corresponding similar star with ten
  photometric measurements, we have to slightly relax the constraints
  to $\Teff(\rm target)\pm100\K$ and [Fe/H]$(\rm target)\pm0.4$ dex.}
The `close' distance is defined in the sense that the star deviates
from the best-fitting $\Teff$--$\CBVint$ line by less than the
uncertainty of 0.06 $\magni$ in the observed color index $(\CBV)_{\rm
  err}$.\footnote{Compared with the IR bands, the intrinsic colors of
  RC stars at optical bands are more sensitive to
  metallicity. Although the optical-to-mid-IR RC $\Teff$--intrinsic
  color relations were all derived in Section 4.1.1, we use the
  $\Teff$--$\CBVint$ relation to discuss the effect of metallicity on
  the RCs' intrinsic color indices.} These stars are considered
reddening-free, since the deviation from the line of intrinsic color
is not larger than the photometric uncertainty. Therefore, the
observed color indices of these stars can represent the intrinsic
colors. This method was used by Yuan et al. (2013) to derive the
empirical extinction law.

Fig~\ref{fig:metallicity}a exhibits the results for the similar stars
selected for the 18 target RC stars in the effective temperature
$\Teff$ vs. observed color index ($B-V$) diagram.
The yellow asterisks represent the 18 target RC stars in the
$l165^{\circ}$ region, the blue solid line denotes the
$\Teff$--$\CBVint$ relation, with the cyan dashed line representing
the 1$\sigma$ envelope. The red asterisks are the average observed
color index values $[\CBV]$ of the similar stars, and these average
values turn out to be the intrinsic color indices of the 18 RC stars.
This method is affected by a systematic bias, as shown in
Figure~\ref{fig:metallicity}a: all reference stars lie below the
best-fitting line, which means that they suffer from some extinction.
Consequently, the derived $(B-V)_0^{\rm metals}$ is redder than the
$(B-V)_0$ from the $\Teff$--$\CBVint$ relation, as shown in
Figure~\ref{fig:metallicity}b. Except for $\CgVint$, this method
results in redder intrinsic color indices $(V-\lambda_x)_0$ in the $r,
i, J, H, \Ks, W1$, and $W2$ bands. The corresponding color excesses
$E(B-V)$ and $E(V-\lambda_x)$ become smaller, and $E(g-V)$ becomes
larger. We also investigated the effect on the $\Rv$ value. The
slightly different intrinsic color indices cause the $\Rv$ value to
become smaller, as shown in Figure~\ref{fig:metallicity}c: $-0.42 <
(\Rv)_{\rm metals} - (\Rv)_0 < -0.14$, while the diversity is still there.

\section{Summary}

The optical-to-mid-IR extinction law has been derived for the diffuse
$l165^{\circ}$ region in the two APASS bands ($B, V$), the three
XSTPS-GAC bands ($g, r, i$), the three {\it 2MASS} bands ($J, H,
\Ks$), and the two {\it WISE} bands ($W1, W2$) using RC stars as
extinction tracers. Specifically, 18 RC stars in this region were
selected from the APOGEE--RC catalog based on their stellar parameters
$\Teff$, $\log g$, and [Fe/H]. The major results of this paper are as
follows:

\begin{enumerate}
\item The stellar intrinsic colors were determined for RC stars with
  effective temperatures in the range $4500\K \le \Teff \le 5100\K$.
  Two methods were adopted, one is based on the analytic
  $\Teff$--intrinsic color relation, the other uses Padova isochrone
  models. The IR intrinsic color indices are consistent with each
  other. Although the optical color indices exhibit notable
  differences, the differences in color are mostly smaller than 0.05
  $\magni$, comparable to the photometric uncertainties.
\item The extinction curves were derived toward sightlines of 18 RC
  stars in the diffuse region around $l=165^{\circ}$. 
  The corresponding $\Rv$ values are determined 
  by fitting the extinction curves with the CCM89 law. 
  The mean $\Rv$ value of 2.8 is consistent with the commonly 
  adopted value for Galactic diffuse clouds ($\Rv =3.1$). 
  However, the $\Rv$ values ranging from 1.7 to 3.8 suggest that 
  the optical extinction law exhibits significant diversity in the 
  $l=165^{\circ}$ region, which is interesting, because it is  
  such a small region (in angular size) of the diffuse ISM. 
  This diversity is beyond the normal expectation that 
  diffuse environments would exhibit an average law with a small variation. 
  Since the extinction law is determined by the dust properties, 
  the result implies that the dust properties are very heterogeneous 
  in their spatial distribution. Consequently, one should be cautious to 
  take an average law to correct for the interstellar extinction. 
  A high spatial resolution study of the extinction law is needed.
\item There is no correlation between $\Rv$ and $E(B-V)$ in the
  $E(B-V)$ range of interest, between 0.2 and 0.6 mag. 
  Since these RC sightlines still coincide with the diffuse region, 
  dense regions are needed to investigate 
  any correlation between $\Rv$ and $E(B-V)$.
\item The distances to our RC sample were derived. They range from
  2.67 kpc to 4.49 kpc. The average visual extinction per kiloparsec,
  $\AV/d$ is 0.37 mag kpc$^{-1}$, which is lower than the average
  value for the Milky Way. There is no apparent relation between $\Rv$
  and the specific visual extinction per kpc.
\end{enumerate}

%%%%%%%%%%%%%%
\acknowledgments{We thank Bingqiu Chen, Jian Gao, and Aigen Li for very helpful discussions, 
  and the anonymous referee for very useful comments/suggestions.
  This research was made possible through the use of AAVSO, 2MASS,
  WISE, LAMOST, and SDSS-III archival data. This work is partially
  supported by the Initiative Postdocs Support Program
  (No. BX201600002), National Natural Science Foundation of China
  projects 11533002, 11503002, 11373015, U1631102, 11373010, and
  11633005, 973 Program 2014CB845702, and the Fundamental Research
  Funds for the Central Universities. S. W. acknowledges support from
  a KIAA Fellowship.}
%%%%%%%%%%%%%%%


\clearpage

\begin{thebibliography}{}
\bibitem[Alam et al.(2015)]{Alam15} Alam, S., Albareti, F. D., Allende
  Prieto, C., et al. 2015, ApJS, 219, 12
\bibitem[Alves(2000)]{Alves00} Alves, D. R. 2000, \apj, 539, 732
\bibitem[Bonatto et al.(2004)]{Bonatto04} Bonatto, C., Bica, E., \&
  Girardi, L. 2004, \aap, 415, 571
\bibitem[Bovy et al.(2014)]{Bovy14} Bovy, J., Nidever, D. L., Rix,
  H.-W., et al. 2014, ApJ, 790, 127
\bibitem[Cabrera-Lavers et al.(2007)]{CL07} Cabrera-Lavers, A.,
  Hammersley, P. L., et al. 2007, \aap, 465, 825
\bibitem[Cardelli et al.(1989)]{CCM89} Cardelli, J. A.,Clayton, G. C.,
  \& Mathis, J. S. 1989, \apj, 345, 245 (CCM89)
\bibitem[Chen et al.(2014)]{Chen14} Chen, B.-Q., Liu, X.-W., Yuan,
  H.-B., et al. 2014, MNRAS, 443, 1192
\bibitem[Chen et al.(2015)]{Chen15}Chen, B.-Q., Liu, X.-W., Yuan,
  H.-B., et al. 2015, MNRAS, 448, 2187
\bibitem[Clark et al.(2012)]{Clark12}Clark, J. S., Najarro, F.,
  Negueruela, I., et al. 2012, \aap, 541, 145
\bibitem[Dame et al.(2001)]{Dame01}Dame, T. M., Hartmann, D., \&
  Thaddeus, P. 2001, ApJ, 547, 792
\bibitem[De\'sertet al.(1988)]{Desert88}De\'sert, F. X., Bazell, D.,
  \& Boulanger, F. 1988. \apj, 334, 815
\bibitem[de Grijs \& Bono(2016)]{deGrijs16} de Grijs, R., \& Bono,
  G. 2016, ApJS, 227, 5
\bibitem[de Vries \& van Dishoeck(1988)]{dv88}de Vries, C. P., \& van
  Dishoeck, E. F. 1988, \aap, 203, 23
\bibitem[Draine(2003)]{Draine03} Draine, B. T. 2003, \araa, 41, 241
\bibitem[Draine(2011)]{Draine11} Draine, B. T. 2011, Physics of the
  Interstellar and Intergalactic Medium (Princeton, NJ: Princeton
  Univ. Press)
\bibitem[Dobashi et al.(2005)]{Dobashi05} Dobashi, K., Uehara, H., et
  al. 2005, \pasj, 57, S1
\bibitem[Ducati et al.(2001)]{Ducati01} Ducati, J. R., Bevilacqua,
  C. M., et al. 2001, \apj, 558, 309
\bibitem[Flaherty et al.(2007)]{Flaherty07} Flaherty, K., Pipher, J.,
  Megeath, S., et al. 2007, \apj, 663, 1069
\bibitem[Gao et al.(2009)]{Gao09} Gao, J., Jiang, B. W., \& Li,
  A. 2009, \apj, 707, 89
\bibitem[Girardi \& Salaris(2001)]{GS01}Girardi, L., \& Salaris, M. 2001, 
 MNRAS, 323, 109
\bibitem[Girardi et al.(2010)]{Girardi10} Girardi, L., Williams,
  B. F., Gilbert, K. M., et al. 2010, ApJ, 724, 1030
\bibitem[Gonz\'alez-Fern\'andez et al.(2014)]{GF14}
  Gonz\'alez-Fern\'andez, C., Asensio Ramos, A., et al. 2014, \apj,
  782, 86
\bibitem[Gottlieb \& Upson(2009)]{GU69} Gottlieb, D., \& Upson,
  W. 1969, ApJ, 157, 611
\bibitem[Groenewegen(2008)]{Groenewegen08} Groenewegen, M. A. T. 2008,
  \aap, 488, 935
\bibitem[Henden \& Munari(2014)]{HM14}Henden, A., \& Munari, U. 2014,
  Contrib. Astron. Obs. Skalnate Pleso, 43, 518
\bibitem[Henden et al.(2016)]{Henden16}Henden, A., Templeton, M.,
  Terrell, D., et al. 2016, VizieR Online Data Catalog, II/336
\bibitem[Hodapp et al.(2004)]{H04} Hodapp, K. W., Kaiser, N., Aussel, H., et al. 2004, AN, 325, 636
\bibitem[Holtzman et al.(2015)]{Holtzman15} Holtzman, J. A., Shetrone,
  M., Johnson, J. A., et al. 2015, AJ, 150, 148
\bibitem[Howell (2011)]{H11} Howell, D. A. 2011, NatCo, 2, 350
\bibitem[Indebetouw et al.(2005)]{Indebetouw05} Indebetouw, R., et
  al. 2005, \apj, 619, 931
\bibitem[Jiang et al.(2006)]{Jiang06} Jiang, B. W., Gao, J., Omont,
  A., Schuller, F., \& Simon, G. 2006, \aap, 446, 551
\bibitem[Liu et al.(2014)]{Liu14} Liu, X.-W., et al., 2014, in
  Feltzing, S., Zhao, G., Walton, N., \& Whitelock, P., eds, Proc. IAU
  Symp. 298, Setting the Scene for Gaia and LAMOST (Cambridge:
  Cambridge Univ. Press), p. 310
\bibitem[Johnson (1966)]{Johnson66} Johnson, H. L. 1966, \araa, 4, 193
\bibitem[Larson et al.(2000)]{Larson00}Larson, K. A., Wolff, M. J., et
  al. 2000, ApJ, 532, 1021
\bibitem[L\'opez-Corredoira et al.(2002)]{LC02} L\'opez-Corredoira,
  M., Cabrera-Lavers, A., et al. 2002, \aap, 394, 883
\bibitem[Lutz et al.(1996)]{Lutz96} Lutz, D., et al. 1996, \aap, 315,
  L269
\bibitem[Lutz et al.(1999)]{Lutz99} Lutz, D. 1999, in The Universe as
  Seen by ISO, eds P. Cox \& M. Kessler (ESA Special Publ., Vol.~427;
  Noordwijk: ESA), p. 623
\bibitem[Marigo et al.(2008)]{Marigo08} Marigo, P., Girardi, L.,
  Bressan, A., et al. 2008, \aap, 482, 883
\bibitem[Mathis(1990)]{Mathis90} Mathis, J. S. 1990, \araa, 28, 37
\bibitem[M\'esz\'aros et al.(2013)]{Meszaros13} M\'esz\'aros, S.,
  Holtzman, J., et al. 2013, \aj, 146, 133
\bibitem[Milne \& Aller(1980)]{MA80} Milne, D. K., \& Aller,
  L. H. 1980, AJ, 85, 17
\bibitem[Munari et al.(2014))]{M14} Munari, U., Henden, A., Frigo, A., 
et al. 2014, AJ, 148, 81
\bibitem[Nataf et al.(2010)]{N10}Nataf, D. M., Udalski, A., Gould, A., et al. 
 2010, ApJ, 721, L28
\bibitem[Nishiyama et al.(2006)]{Nishiyama06} Nishiyama, S., Nagata,
  T., Kusakabe, N., et al. 2006, \apj, 638, 839
\bibitem[Nishiyama et al.(2009)]{Nishiyama09} Nishiyama, S., Tamura,
  M., Hatano, H., et al. 2009, \apj, 696, 1407
\bibitem[Planck Collaboration XIII (2014)]{Planck14} Planck
  Collaboration XIII, 2014, \aap, 571, A13
\bibitem[Sarajedini(1990)]{S99} Sarajedini, A. 1999, AJ, 118, 2321
\bibitem[Schlafly \& Finkbeiner(2011)]{SF11}Schlafly, E. F., 
  \& Finkbeiner, D. P. 2011, ApJ, 737, 103
\bibitem[Schlafly et al.(2016)]{S16}Schlafly, E. F., Meisner, A. M., 
 Stutz, A. M., et al. 2016, \apj, 821, 78
\bibitem[Schultheis et al.(2014)]{Schultheis14} Schultheis, M.,
  Zasowski, G., et al. 2014, \aj, 148, 24
\bibitem[Skrutskie et al.(1997)]{Skrutskie97}Skrutskie, M. F., et al.,
  1997, in Garzon, F., Epchtein, N., Omont, A., Burton, B., \& Persi,
  P., eds, Astrophys. Space Sci. Libr., 210, The Impact of Large Scale
  Near-IR Sky Surveys (Dordrecht: Kluwer), p. 25
\bibitem[Skrutskie et al.(2006)]{Skrutskie06} Skrutskie, M. F., et
  al. 2006, \aj, 131, 1163
\bibitem[Snow \& McCall(2006)]{SM06} Snow, T. P., \& McCall,
  B. J. 2006, \araa, 44, 367
\bibitem[Torres et al.(1991)]{Torres91} Torres-Dodgen, A. V., Carroll,
  M., \& Tapia, M. 1991, MNRAS, 249, 1
\bibitem[Wainscoat et al.(1992)]{Wainscoat92} Wainscoat, R., Cohen,
  M., Volk, K., et al. 1992, \apjs, 83, 146
\bibitem[Wang et al.(2013)]{Wang13} Wang, S., Gao, J., Jiang, B. W.,
  Li, A., \& Chen, Y. 2013, \apj, 773, 30
\bibitem[Wang \& Jiang(2014)]{WJ14} Wang, S., \& Jiang, B. W. 2014,
  \apjl, 788, L12
\bibitem[Welty \& Fowler(1992)]{WF92} Welty, D. E., \& Fowler, J. R. 
 1992, ApJ, 393, 193
\bibitem[Wright et al.(2010)]{}Wright, E. L. et al. 2010, AJ, 140,
  1868
\bibitem[Xue et al.(2016)]{Xue16} Xue, M., Jiang, B. W., Gao, J., et
  al. 2016, \apjs, 224, 23
\bibitem[Yuan et al.(2013)]{Yuan13} Yuan, H. B., Liu, X. W., \& Xiang,
  M. S. 2013, MNRAS, 430, 2188
\bibitem[Zasowski et al.(2009)]{Zasowski09} Zasowski, G., Majewski,
  S. R., Indebetouw, R., et al. 2009, \apj, 707, 510
\bibitem[Zhang et al.(2014)]{}Zhang, H. H., Liu, X. W., Yuan, H. B.,
  et al. 2014, Res. Astron. Astrophys., 14, 456


\end{thebibliography}
\end{document}